\RequirePackage{rotating}
\documentclass[useAMS,usenatbib]{mnras}
\pdfoutput=1
\usepackage{psfig}
\usepackage{amsmath}
\usepackage{amssymb}
\usepackage{upgreek}
\usepackage{longtable}
\usepackage[mathscr]{eucal}
\usepackage{appendix}
\usepackage{graphicx}
\usepackage{float}
\usepackage{afterpage}
\usepackage{subcaption}
\usepackage[export]{adjustbox}
\usepackage[usenames, dvipsnames]{color}
\usepackage{titlesec}

\titlespacing*{\section}{0pt}{1.1\baselineskip}{\baselineskip}
\titlespacing*{\subsection}{0pt}{1.1\baselineskip}{\baselineskip}

\title[Low-mass Orion Nebula Cluster Runaways]{Low-mass Runaways from the Orion Nebula Cluster - Kinematic Age Constraints on Star Cluster Formation}
\author
[Fajrin et al.]
{Muhammad Fajrin$^{1,2}$, 
Joseph J. Armstrong$^{2}$, 
Jonathan C. Tan$^{2,3}$, 
Juan P. Farias$^4$, 
\newauthor and Laurent Eyer$^{5}$\\
$^{1}$Astronomy Study Program, Faculty of Mathematics and Natural Sciences, Institut Teknologi Bandung,\\ Jalan Ganesha 10, Bandung 40132, Indonesia   \\
$^{2}$Department of Space, Earth \& Environment, Chalmers University of Technology, SE-412 96 Gothenburg, Sweden \\
$^{3}$Department of Astronomy, University of Virginia, Charlottesville, VA, USA\\
$^{4}$Department of Astronomy, University of Texas, Austin, TX, USA\\
$^{5}$Department of Astronomy, University of Geneva, Chemin Pegasi 51, CH-1290, Versoix, Switzerland \\
}
\date{Accepted XXX. Received YYY; in original form ZZZ}
\pubyear{2024}

\begin{document}
\label{firstpage}
\pagerange{\pageref{firstpage}--\pageref{lastpage}}
\maketitle

\begin{abstract}
In their early, formative stages star clusters can undergo rapid dynamical evolution leading to strong gravitational interactions and ejection of ``runaway'' stars at high velocities. While O/B runaway stars have been well studied, lower-mass runaways are so far very poorly characterised, even though they are expected to be much more common. We carried out spectroscopic observations with MAG2-MIKE to follow-up 27 high priority candidate runaways consistent with having been ejected from the Orion Nebula Cluster (ONC) $>2.5$ Myr ago, based on Gaia astrometry.
%\citep{farias20}. 
We derive spectroscopic youth indicators (Li \& H$\alpha$) and radial velocities, enabling detection of bona fide runaway stars via signatures of youth and 3D traceback.
%\citep[e.g,][]{armstrong22}. 
We successfully confirmed 11 of the candidates as low-mass Young Stellar Objects (YSOs) on the basis of our spectroscopic criteria and derived radial velocities (RVs) with which we performed 3D traceback analysis. Three of these confirmed YSOs have kinematic ejection ages $>4\:$Myr, with the oldest being 4.7~Myr.
%are consistent with having being ejected from the ONC and have 3D traceback timescales up to 4.5 Myr, 
Assuming that these stars indeed formed in the ONC and were then ejected, this yields an estimate for the overall formation time of the ONC to be at least $\sim 5\:$Myr, i.e., about 10 free-fall times, and with a mean star formation efficiency per free-fall time of $\bar{\epsilon}_{\rm ff}\lesssim0.05$. These results favor a scenario of slow, quasi-equilibrium star cluster formation, regulated by magnetic fields and/or protostellar outflow feedback.
%Even one single confirmed runaway would represent a breakthrough in extending the age estimate of the ONC via the ejection age method beyond the 2.5~Myr set by $\mu$ Col \& AE Aur (Hoogerwerf et al. 2001, A\&A, 365, 49), yielding crucial constraints on cluster formation models (Tan et al. 2006, ApJ Lett., 641, 121; Farias et al. 2019, MNRAS, 483, 4999). For this pilot project, we have selected the 25 highest priority targets from the Farias et al. (2020) candidate list, i.e., being relatively bright and with 2D traceback ages $>2.5\:$Myr.
\end{abstract}
	
	
\section{Introduction}

Stars tend to form in clusters from dense gas clumps within giant molecular clouds (GMCs) 
%with a significantly greater concentration of stars per volume than the surrounding field
\citep{lada03}. In their early, gas-dominated stages they may undergo significant dynamical evolution which can lead to regions of enhanced stellar densities, mass segregation, processing of multiple systems, and ejection of ``runaway'' or ``walkaway'' stars \citep[e.g.,][]{marks12,parker14,farias19}.
%change their stellar density and internal structure, and result in dynamical mass segregation and the destruction of multiple systems \citep{marks12,parker14}. It is thus expected that stars within clusters have significant probabilities of undergoing strong dynamical interactions that can lead to their ejection at high velocities \citep{farias19}. 
The kinematic ``ejection age'' of such stars can provide an important constraint on the age of a cluster, which is independent and complementary to ages based on pre-main sequence stellar evolutionary models.
%a unique insight into the history of young clusters and star forming regions. 
In particular, the oldest ejected runaways from a cluster offer model independent lower limits on cluster age. For a still forming cluster, the cluster age gives a lower limit on the age spread of the system and thus an upper limit on the time averaged star formation rate (SFR), or equivalently the star formation efficiency per free-fall time ($\bar{\epsilon}_{\rm ff}$).
This is a basic parameter which can help distinguish different theoretical models of star cluster formation, i.e., between those involving ``fast'' formation within one or a few free-fall times \citep[e.g.,][]{elmegreen00} and those assuming ``slow'', quasi-equilibrium formation \citep[e.g.,][]{tan06}.
%for cluster formation theories \citep[e.g.,][]{tan06} upon which many star formation properties depend, such as efficiency per free-fall time \citep{dario14} and massive star formation timescale \citep{wang10}. 
Furthermore, the fraction of stars that become runaways depends sensitively to the duration of the dense, early, gas-rich phase \citep{oh16,farias19}. So an accurate assessment of this timescale from finding oldest known runaways enables a more accurate prediction of the global runaway population.

%frequency of the runaway star population are greatly dependent on the densities \citep{oh16} and lifetimes \citep{farias19} of the early cluster environments. 
%Therefore, characterizing the unbound population provides valuable constraints on the dynamical history of forming stellar clusters. 
% Joe - when giving citations try to avoid having brackets within brackets, and I'm not sure you need to have "(e.g., citation)", but just "citation" is fine, even if the citations you give are only an example out of many papers that support your point. 
%--

\begin{figure*}
    \centering
    \includegraphics[width=1\textwidth]{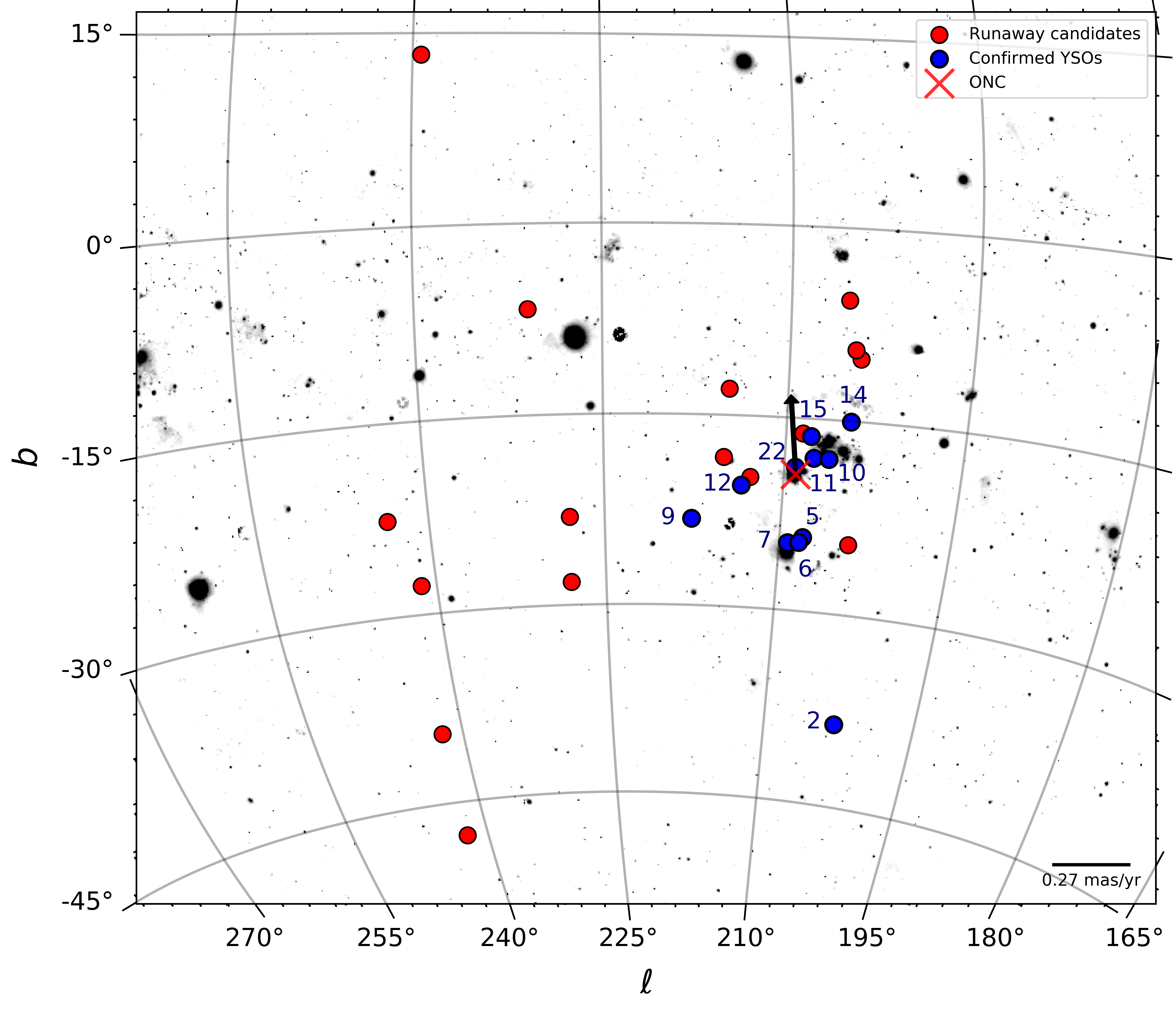}
    \caption{Spatial distribution of the 27 candidate runaways for spectroscopic follow-up. Those candidates with detected Li absorption, i.e., most likely to be YSOs, are colored blue. The 'x' symbol indicates the location of the ONC. The arrow indicates the proper motion of the ONC and the magnitude scale (mas/yr) of proper motion vector is indicated by the scale bar in the bottom right. The background is DSS2 Blue map of region around the ONC accessed from ALADIN.} %{\bf LaurentC1: I count only 26 points in the figure} }
    \label{fig:sky}
\end{figure*}
% Joe - perhaps there is a way to make the targets stand out more in this plot, perhaps have the candidates as solid red dots and then those with Li as solid blue, it's difficult to tell them apart from size alone.

Most known runaway stars are bright O- and B-type stars (e.g., \citealt{tetzlaff11}), since they are easier to observe than fainter, lower-mass stars. However, N-body simulations \citep[e.g.,][]{schoettler19,farias19} predict that most runaway stars will be of low mass.
%stars in a wide mass range 
%(down to 0.1$M_\odot$) 
%will also be ejected from young star forming regions. 
With the availability of high-precision 5-parameter astrometry from Gaia \citep{gaia20a}, we are now able to extend our search for runaways to these numerous low-mass stars too. However, with only plane of sky proper motion and spatial information that enables a ``2D-traceback'' type analysis, there are generally many field star interlopers that can masquerade as runaway candidates \citep[e.g.,][]{farias20}. Radial velocity information can help reduce this contamination,
%jct - I don't think RV is the limiting factor.
but current Gaia releases do not provide radial velocities (RVs) for relatively faint stars, and those that are available have large uncertainties ($\sim$15 km/s for sources with V = 15.7~mag). To analyse the full 3D velocities of low-mass candidate runaways, it is necessary to combine Gaia astrometry with spectroscopic radial velocities. The same spectroscopic observations can also yield indicators of stellar youth, e.g., H$\alpha$ emission or Li absorption, which are then the key tests for secure identification of a low-mass runaway star from a young cluster.

% Joe - The point about the Gaia RV precision being 15 km/s at Vmag = 15.7 needs a reference, does it come from a Gaia Collaboration paper? 
%--

The Orion Nebula Cluster (ONC) is the nearest dense \citep[$>10^4 M_\odot pc^{-3}$; see ][]{hillenbrand97} cluster that is still forming stars ($\sim$400 pc; e.g., \citealt{kuhn19}, $\sim$4 Myr; e.g., \citealt{dario16}), making it an important test case for theories of star cluster formation. Recent studies have identified high velocity stars consistent with having been ejected from the ONC \citep{mcbride19,schoettler20,farias20,platais20,bobylev21}, but these works lacked precise radial velocities for the majority of their runaway candidates and so have been largely limited to analysis in 2D. In particular, \citet{farias20} used Gaia DR2 proper motions to search for runaway candidates in a 45$^{\circ}$ radius around the ONC, in combination with Gaia and WISE photometric classifications \citep{marton19} and optical variability \citep{cody}, to identify young stellar objects (YSOs) consistent with the age of the ONC. Using their best candidates, \citet{farias20} constructed a high velocity distribution for the ONC that was compared with N-body simulations, showing that the dynamical history of the ONC is consistent with a dense primordial environment (with mass surface densities of $\sim$1 g/cm$^2$) and low star formation efficiency per free fall time ($\sim$1\%).  However, only $7\%$ of their proper motion candidates had measured radial velocities with which full 3D traceback could be calculated. Fewer than a third of sources with radial velocities had 3D traceback ages within 1 Myr of their 2D traceback ages, further highlighting the need for precise radial velocities. Therefore, it is imperative that the estimated high-velocity distribution of the ONC is cleaned of contaminants in order to determine the best cluster formation models that can reproduce it.  

\begin{table*}
\begin{center}
{\renewcommand{\arraystretch}{1.5}
\begin{tabular}
{|p{1.0cm}|p{2.8cm}|p{1.1cm}|p{1.1cm}|p{0.6cm}|p{0.4cm}|p{2.3cm}|p{0.5cm}|p{1.2cm}|p{0.6cm}|p{1.1cm}|p{1.1cm}| }
\hline
ID & Gaia ID & $\ell$ ($^{\circ}$) & $b$ ($^{\circ}$) & Gmag & Gaia YSO & $v_r$ \newline (km s$^{-1}$) & score & $v_{t0}$ \newline (km s$^{-1}$) & CA (') & $t_{\rm 3D,opt}$ (Myr) & Exposure time (s) \\
\hline
OBJ-1 & 4844159373956999936 & 242.3397 & -47.6840 & 15.26 & & & bl*! & 37.58 & 15.76 & 4.13$_{-1.86}^{+2.64}$ & 3$\times250$ \\
OBJ-2 & 3192134597649605376 & 203.0002 & -38.5044 & 16.47 & \checkmark & & al & 21.01 & 8.64 & 5.14$_{-2.12}^{+0.85}$ & 3$\times500$ \\
OBJ-3 & 4817925576274600192 & 242.0673 & -39.4966 & 16.26 & & & bl*! & 63.10 & 23.92 & 3.82$_{-1.79}^{+0.62}$ & 3$\times450$ \\
OBJ-4 & 3212944607551376384 & 203.9799 & -24.5334 & 16.99 &  & & bl! & 17.08 & 0.17 & 3.05$_{-1.79}^{+1.31}$ & 3$\times600$ \\
OBJ-5 & 3207501131641282176 & 208.2989 & -24.6643 & 16.52 &  & & al & 5.57 & 10.06 & 5.23$_{-2.57}^{+6.15}$ & 3$\times550$ \\
OBJ-6 & 3208291783581908608 & 208.0032 & -24.2580 & 16.89 & \checkmark & 32.485 ± 3.562$^{(a)}$ & bl & 5.62 & 9.81 & 4.56$_{-1.21}^{+5.30}$ & 3$\times600$ \\
OBJ-7 & 3207022053810350976 & 209.2567 & -24.7423 & 16.68 & \checkmark  & & al & 6.68 & 12.82 & 4.19$_{-1.95}^{+1.39}$ & 3$\times550$ \\
OBJ-8 & 2957497235735128448 & 228.0767 & -28.1871 & 16.99 & & & bl* & 45.03 & 5.43 & 2.43$_{-1.17}^{+1.51}$ & 3$\times600$ \\
OBJ-9 & 2984454031031531008 & 217.6080 & -23.1807 & 15.40 & & & al*! & 8.07 & 16.43 & 5.57$_{-2.79}^{+0.73}$ & 3$\times250$ \\
OBJ-10 & 3216889827071056896 & 206.3162 & -18.0608 & 16.46 & \checkmark  & & bl & 12.19 & 6.71 & 1.68$_{-0.94}^{+1.29}$ & 3$\times500$ \\
OBJ-11 & 3215804677813294976 & 207.5906 & -18.0482 & 16.81 & \checkmark  & & bl & 8.53 & 9.10 & 1.42$_{-0.78}^{+1.38}$ & 3$\times600$ \\
OBJ-12 & 3012142379518284288 & 213.5078 & -20.4489 & 15.37 &  & & bl* & 9.77 & 4.56 & 2.56$_{-1.37}^{+1.20}$ & 3$\times250$ \\
OBJ-13 & 3015308629408804608 & 212.7895 & -19.7749 & 14.51 & & $39.946\pm 0.702^{(b)}$ & bl*! & 8.35 & 0.13 & 2.88$_{-1.89}^{+0.79}$ & 3$\times150$ \\
OBJ-14 & 3219378365481960832 & 204.7270 & -15.0406 & 14.96 & &  & bl & 8.60 & 13.72 & 4.12$_{-2.04}^{+2.39}$ & 3$\times150$ \\
OBJ-15 & 3216174629116142336 & 207.9220 & -16.3723 & 14.39 & \checkmark &  & bl & 10.93 & 13.28 & 1.74$_{-0.78}^{+1.92}$ & 3$\times100$ \\
OBJ-16 & 2888109908763598976 & 241.3582 & -27.7466 & 15.39 &  &  & al*! & 26.29 & 8.97 & 4.79$_{-1.49}^{+2.81}$ & 3$\times250$ \\
OBJ-17 & 3011187006993509504 & 215.0589 & -18.3087 & 15.22 &  &  & bl*! & 26.98 & 9.18 & 1.67$_{-1.21}^{+1.23}$ & 3$\times250$ \\
OBJ-18 & 3316420643274767488 & 204.2695 & -10.1560 & 15.65 &  &  & bl! & 25.87 & 20.00 & 3.81$_{-2.46}^{+1.51}$ & 3$\times350$ \\
OBJ-19 & 3315632671394273024 & 204.7219 & -9.4776 & 13.48 & & $107.556 \pm 1.452^{(a)}$ & bll & 33.08 & 18.13 & 2.19$_{-1.27}^{+1.16}$ & 3$\times80$ \\
OBJ-20 & 3018141830356350976 & 214.7763 & -12.9386 & 16.31 &  &  & bl! & 30.53 & 18.15 & 2.14$_{-1.34}^{+1.77}$ & 3$\times450$ \\
OBJ-21 & 2932903703242234112 & 230.9026 & -6.7062 & 12.30 &  & $87.791 \pm 1.043^{(a)}$ & bll*! & 21.96 & 34.66 & 4.12$_{-3.68}^{+2.40}$ & 3$\times50$ \\
OBJ-22 & 3017382033474172800 & 209.0747 & -18.8395 & 14.18 & \checkmark &  & bl & 4.47 & 2.22 & 1.01$_{-0.69}^{+2.40}$ & 3$\times100$ \\
OBJ-23 & 3023944262453551232 & 208.6032 & -16.1726 & 14.90 &  &  & bl & 11.27 & 10.71 & 1.91$_{-0.99}^{+2.16}$ & 3$\times150$ \\
OBJ-24 & 2915784994393795456 & 227.9664 & -23.0595 & 15.66 &  &  & bl*! & 49.27 & 17.32 & 3.20$_{-1.73}^{+1.50}$ & 3$\times350$ \\
OBJ-25 & 2885209740687428224 & 243.5372 & -22.5201 & 15.46 &  &  & bl*! & 42.56 & 9.85 & 3.73$_{-1.20}^{+2.15}$ & 3$\times300$ \\
OBJ-26 & 3317517165606496256 & 205.4203 & -5.6509 & 16.01 &  &  & bl! & 38.11 & 4.14 & 2.88$_{-1.59}^{+1.78}$ & 3$\times450$ \\
OBJ-27 & 5709959085012172928 & 239.3071 & 13.3290 & 16.43 &  & $79.343 \pm 1.249^{(a)}$ & bl! & 66.91 & 19.63 & 3.45$_{-1.58}^{+1.70}$ & 3$\times500$ \\

\hline
\end{tabular}}
\end{center}
\setlength{\belowcaptionskip}{-10pt}
\setlength{\textfloatsep}{0pt}
\caption{List of candidate runaways observed with MAG2-MIKE. Columns are: Target identifier in the observation, Gaia DR3 ID number, Galactic longitude (Gaia DR3), Galactic latitude (Gaia DR3), Gaia DR3 \textit{G}mag, Gaia YSO flag \protect\citep{rimoldini23, marton23}, $v_r$ from (a) Gaia, and (b) APOGEE taken from \protect\cite{tsantaki22}, \protect\citet{farias20} score for youth criteria met (``a'' for sources that pass the YSOflag criteria, ``b'' for sources that only fail one of either YSOflag, WYSOflag or VARflag criteria, ``I'' for sources that pass both WYSOflag and VARflag criteria, ``II'' for sources that pass either WYSOflag or VARflag criteria, ``*'' for sources whose 2D closest approach overlaps with the center of the ONC within 1$\sigma$ uncertainty, ``!'' for sources which fail the RVflag criteria), 2D ejection velocity \protect\citep{farias20}, 2D closest approach to the ONC \protect\citep{farias20}, 'optimal' 3D traceback time \protect\citep{farias20}, exposure time per source.}
\label{targets_table}
\end{table*}

Even one single confirmed runaway would represent a breakthrough in extending the age estimate of the ONC via the ejection age method beyond the $\sim$2.5 Myr set by $\mu$ Col \& AE Aur \citep{hoogerwerf01}, yielding crucial constraints on cluster formation models \citep{tan06,farias19}. For this project, we have selected the 27 highest priority targets from the \citet{farias20} candidate list updated with Gaia EDR3 astrometry, i.e., being relatively bright and with 2D traceback ages $>2.5$ Myr. We have carried out spectroscopic observations with MAG2-MIKE to follow-up these candidate runaways
%consistent with having been ejected from the Orion Nebula Cluster (ONC) $>2.5$ Myr ago, based on Gaia astrometry \citep{farias20}, 
in order to confirm their youth with spectroscopic indicators (i.e., Li and H$\alpha$) and to derive radial velocities to enable 3D-traceback 
%of their past trajectories 
to determine the likelihood of their origin in the ONC and the time of their ejection.
 
\section{Observational Methods}
\label{section_data}

\begin{figure*}
    \centering
	\includegraphics[width=0.7\textwidth]{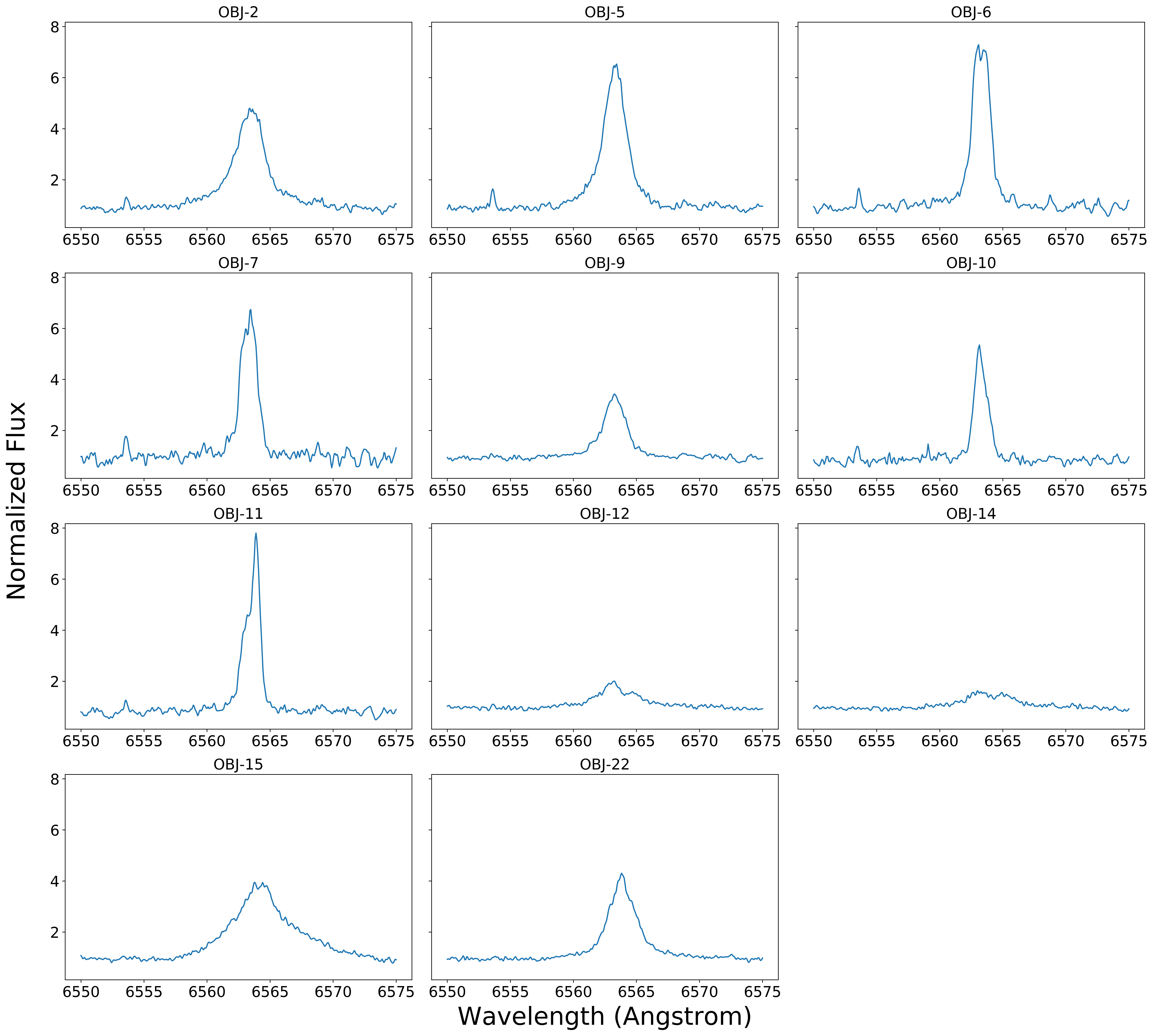}
    \includegraphics[width=0.7\textwidth]{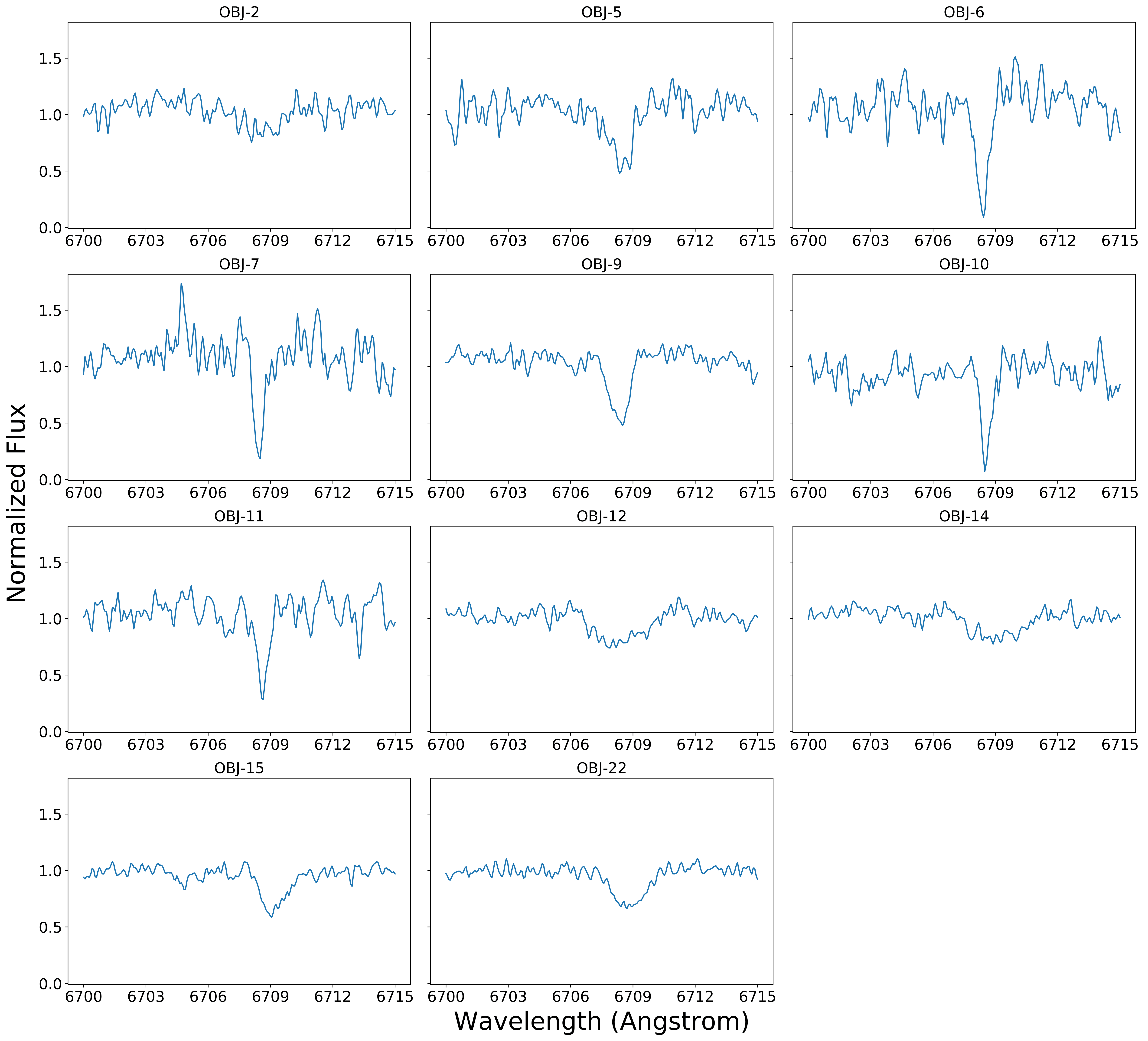}
    \caption{H$\alpha$ emission (\textit{top)} and lithium 6708 \AA \ (\textit{bottom}) profiles in combined spectra of selected candidates. The fluxes are normalized non-calibrated counts.}
    \label{fig:Ha}
\end{figure*}

\subsection{Target selection}

To select targets for spectroscopic follow-up, we updated the \citet{farias20} list of 16 994 candidate runaways with Gaia EDR3 astrometry, which has typical improvements in precision of proper motion by 33\% and parallax by 50\% compared to Gaia DR2 \citep{lindegren21}. We then re-calculated traceback parameters, such as 2D (plane of sky) closest approach to the ONC, ejection velocity in 2D, and traceback time to closest approach for candidates with reliable Gaia EDR3 astrometry (RUWE $<$ 1.4). 
% Joe - the precision improvement % needs a reference too, it likely comes from the Gaia Collaboration EDR3 astrometry paper - first author Lindegren I think
%--

We cross-matched our candidate runaway list with the Gaia DR3 variable YSO catalogue \citep{rimoldini23, marton23} to use this as another youth indicator. We also cross-matched the sample with the radial velocity compilation from \textit{Survey of Surveys} \citep{tsantaki22} to recalculate the 3D traceback for those targets with known RVs. 

% {\bf LaurentC2 comment out of the 10 stars, 6 are classified as YSO from the ML technics from the light curve behaviour published in DR4. I wonder why there is the different with Gabor Marton's article.} 
% Joe - these data should also be included in Table 1., a column for Gaia DR3 YSO variability, a column for literature RV and a column for which survey or paper the RV comes from. 
%--

Runaway candidates were selected for spectroscopic observations if they passed two or more youth criteria (YSOflag, WYSOflag, VARflag or Gaia DR3 variable YSO match, see \citealt{farias20}), had an 84th percentile predicted 3D traceback time ($t_{3D\ opt}$) greater than 2.5 Myr, a 2D closest approach consistent with originating within the cluster radius \citep[10 arcmin corresponding to 1.2 pc,][]{farias20}, an ejection velocity ($v_{t0}$) greater than $4\:{\rm km\:s}^{-1}$, and an estimated mass / $T_{\rm eff}$ (based on position in a Gaia colour-magnitude diagram) consistent with being in the mass range where Li is an effective youth indicator \citep[e.g.,][]{soderblom10}. In the 2D traceback calculations we account for the peculiar motion of the Sun using velocities from \citet{schonrich10}. In total this gives us 27 candidate runaways for spectroscopic follow-up (Table~\ref{targets_table}). In Fig.~\ref{fig:sky}, we illustrate how the candidates are spread across the sky. We will refer the candidates based on their identifiers in this table for ease of reference.

\subsection{Observations}
\label{section_observations}

Observations took place on the 16th December 2022 and 22nd February 2023 using the Magellan Inamori Kyocera Echelle (MIKE) spectrograph on Magellan-Clay 2 at Las Campanas Observatory (LCO). The 1.0\arcsec slit with 2 $\times$ 2 binning was used, yielding $R\sim$\ 22,000 in the red and $R\sim$\ 28,000 in the blue, respectively. All spectra cover the wavelength range from $\sim$3860 \AA\ to $\sim$9000 \AA. Exposure times were estimated for each star using the LCO exposure time calculator to achieve a combined SNR $>$ 20 from 3 exposures, allowing us to measure RVs and equivalent widths of Li and H$\alpha$. 

For each target, Th-Ar lamp exposures were taken as well as a set of 10 milky flats at the beginning of the night. Targets were observed in a slit pair mode, where that target spectrum is observed in one slit while a sky spectrum is observed in the other. Between multiple exposures the slits used for the target and the sky spectrum are alternated.

\subsection{Data Reduction}
\label{section_reduction}

The spectroscopic data were reduced according to standard procedures using \texttt{IRAF}. The processes include data cleaning (flat-fielding, cosmic ray removal, and sky subtraction), spectral/aperture extraction, and wavelength calibration using Th-Ar comparison spectra. The reduction process resulted in multi-order spectra, which then were merged into a single spectrum and normalised using \texttt{IRAF} task \texttt{continuum} and \texttt{scombine}.

% Joe - check that the dashed line is in the right place, it looks like its at -0.2 rather than 0.2
% Joe - check that the text and figure have the right colour, is it J-W4 or K-W4?

\section{Results}
\label{section_spectroscopy}

\subsection{Signatures of Youth}

The youth signatures for the targets in general are predefined in Table~\ref{targets_table} under the column labeled ``score'', which are described in the caption \citep[see][for further details about how these were derived]{farias20}. In Table~\ref{targets_table} we can see that all the targets either satisfy all the youth signatures or only fail in one. Also, this ``failure'' can be due to the source not being able to be evaluated in this metric, such as the lack of IR photometry needed to evaluate the WYSOflag for ~75\% of all candidates \citep{farias20}. Therefore, we expect that the selected targets do already have a high likelihood of being YSOs. Here we report on their additional youth indicators of Li and H$\alpha$ (Fig.~\ref{fig:Ha}) and then further examine their Gaia variability properties and their degree of IR excess.

Since we do not have multi-epoch spectra for our targets, we cannot rule out the possibility that some may be spectroscopic binaries. However, runaway stars ejected by dynamical disruption of multiple systems are expected to be predominantly single stars \citep{leonard90,perets12,schoettler22}. It is also unlikely that a spectroscopic binary system will have an observed radial velocity which happens by chance to be consistent with ejection from the ONC. Therefore, we continue our analysis assuming that these are single stars.

% Joe - the Y axis label should then be "normalized flux"

\begin{figure}
    \centering
    \includegraphics[width=0.5\textwidth]{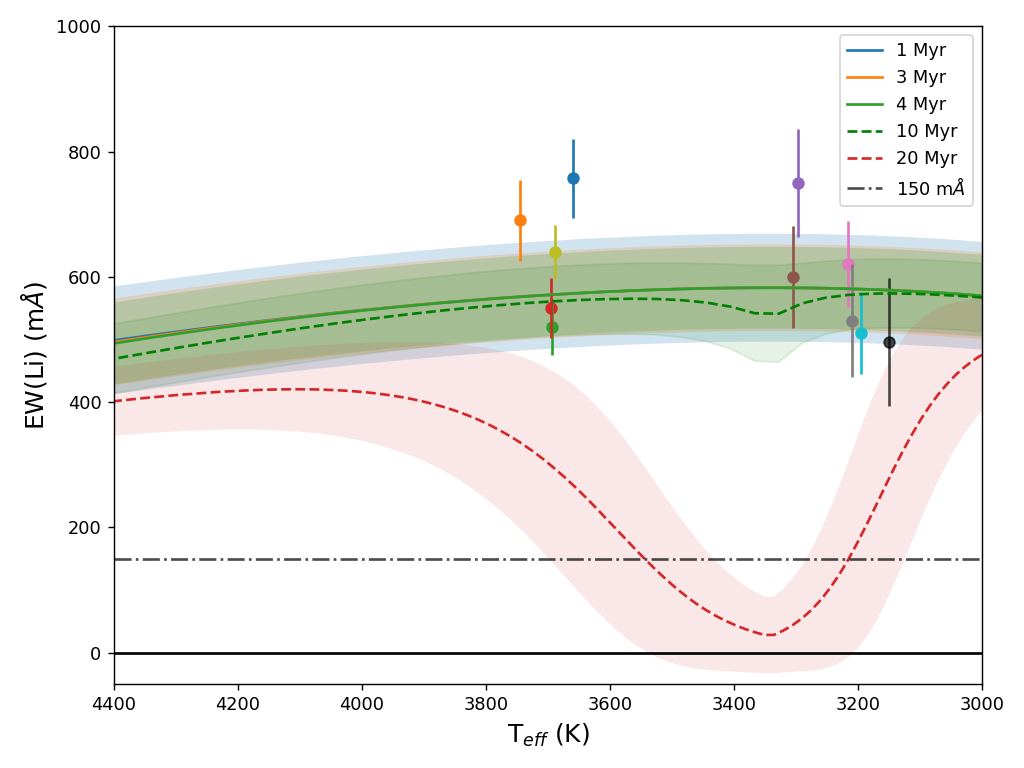}
    \caption{Effective temperature ($T_{\rm eff}$) versus equivalent-width of Li (EW(Li)) for 11 runaway candidates that meet our spectroscopic youth criteria. Coloured lines are EAGLES Li depletion models \protect\citep{jeffries23} for 1, 3, 4, 10 and 20 Myrs, with the shaded regions indicating their 1$\sigma$ uncertainty. $T_{\rm eff}$s were estimated by comparison to PARSEC \protect\citep{marigo17} stellar evolution models on a de-reddened Gaia DR3 BP-RP v G-RP colour-colour diagram (see Section \ref{Liequivalentwidths}).} %{\bf LaurentC5: Where is the figure 4 referenced in the text?}}
    \label{fig:li_teff}
\end{figure}

\subsubsection{Li \& H$\alpha$ Equivalent Widths}
\label{Liequivalentwidths}
Stars with high levels of magnetic activity (and therefore young) should exhibit hydrogen emission features. The youngest stars may also have ongoing accretion. The presence of lithium in the photospheres of low-mass stars can also be used to identify young stars \citep{soderblom10}. Low-mass, fully convective stars are particularly efficient at burning lithium, which would then no longer be visible in the photosphere after a certain time. If the EW of the lithium 6708 \AA\ line in such stars is several hundred m\AA\, the star is likely to be younger than 20-30 Myr.

% Joe - references needed here, e.g. Soderblom et al. 2010 for the explanation of Li as a youth indicator. 
%Also, it should be noted that the rate of Li depletion changes with Teff, as the interior structure changes. So for PMS stars with Teff ~3300K, their Li depletes significantly in ~15 Myr, while slightly hotter PMS stars ~4500K can still exhibit EW(Li)> 0.2 A for up to about 50 Myrs. See Fig. 2 of Jeffries et al. 2023 for an illustration of this.

We therefore look for the presence of H$\alpha$ and Li 6708~\AA\ in the spectra of the observed candidates. We measure the equivalent widths of H$\alpha$ and Li 6708 \AA\ using \texttt{IRAF} task \texttt{splot}, where we fit the lines with Gaussian profiles. The measurements are done for the individual exposures of each source and we then calculate the mean EW value for every star. Uncertainties were calculated using the Cayrel formula \citep{cayrel88}, which assumes a Gaussian line profile and depends on the full width at half-maximun of the line, on the pixel size (in wavelength units) and on the S/N ratio. Measured equivalent-widths and their respective uncertainties are given in Table \ref{results_table}.

%jct - reorganise the text here... first present the Li results in one paragraph, then the H-alpha in another.

Out of 27 observed targets, we found 11 stars with EW(Li) above the thresholds commonly used for YSO signatures \citep[e.g., EW(Li) $>$ 0.1-0.2 \AA\ ][]{jeffries14,armstrong20,armstrong22}. EW(Li) for each of these 11 targets is notably large, ranging from 0.44 \AA\ for the smallest to 0.81 \AA\ for the largest. In Figure~\ref{fig:li_teff}, we plotted the effective temperature (estimated by comparing position in a de-reddened Gaia DR3 BP-RP vs G-RP colour-colour diagram with PARSEC \citet{marigo17} stellar evolution models, using extinction and reddening estimates for each source, taken from \citealt{farias20}.) versus EW(Li) for these targets. And based on comparison to the EAGLES Lithium depletion model \citep{jeffries23}, our targets are consistent with ages $<10\:$Myr.

We also found 5 stars that exhibit EW(H$\alpha$) above the thresholds commonly used for YSO signatures \citep[e.g., EW(H$\alpha$) $>$ 10 \AA,][]{nikoghosyan19,armstrong22}, which are all among the candidates with high EW(Li)s. In the 11 Li-rich candidates, EW(H$\alpha$) varies from a minimum of 4.56 \AA\ to a maximum of 18.61 \AA. 

In Fig.~\ref{fig:Ha}, we present the detected H$ \alpha$ emission line and Li 6708 \AA\ profile. In conclusion, we confirm that 11 out of 27 targets are YSOs based on the presence of 
%H$\alpha$ and 
lithium in their optical spectra. In addition half of these YSOs have H$\alpha$ emission, which may indicate accretion activity.

%It is also worth noting that none of these YSOs show doubled or blended lines in their spectra, so there is no indication that any of these are spectroscopic binaries. 

\subsubsection{Variability}

%jct - summarise all 27 sources in terms of the Farias et al. variability measure and the latest DR3 variability results.
Out of the 27 observed runaway candidates 12 meet the VARflag criteria of \citet{farias20}: OBJ-2, 5, 7, 9, 12, 14, 15, 16, 19, 21, 22 \& 23. Out of these, however, OBJ-16, 19, 21 \& 23 were not also confirmed as YSOs by EW(Li).

Out of the 27 observed runaway candidates we found 7 that were included in the Gaia DR3 YSO variability catalogue \citep[see ~\ref{targets_table}][]{rimoldini23, marton23}, all of which were also confirmed as a YSO by EW(Li).

Interestingly, only 4 candidate runaways both meet the \citet{farias20} VARflag criteria and are included in the Gaia DR3 YSO variability catalogue, OBJ-2, 7, 15 \& 22, all 4 of which are confirmed as YSOs by EW(Li). 

\begin{figure}
    \centering    
    \includegraphics[width=0.5\textwidth,center]{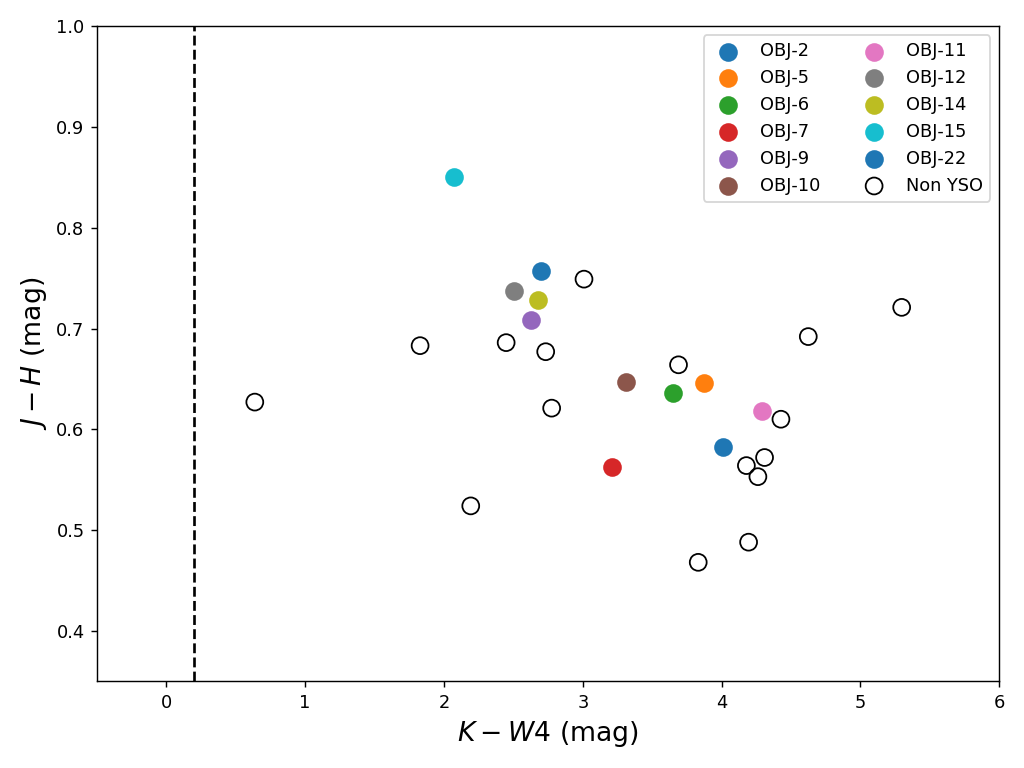}
    \caption{Diagram of \textit{J} $-$ \textit{H} vs. \textit{K} $-$ \textit{W4} of the runaway candidates, where coloured points indicate our 11 confirmed YSOs. The black dashed line illustrates the threshold for IR excess.}
    \label{fig:ir_excess_wu}
\end{figure}

\begin{figure}
	\includegraphics[width=0.5\textwidth,center]{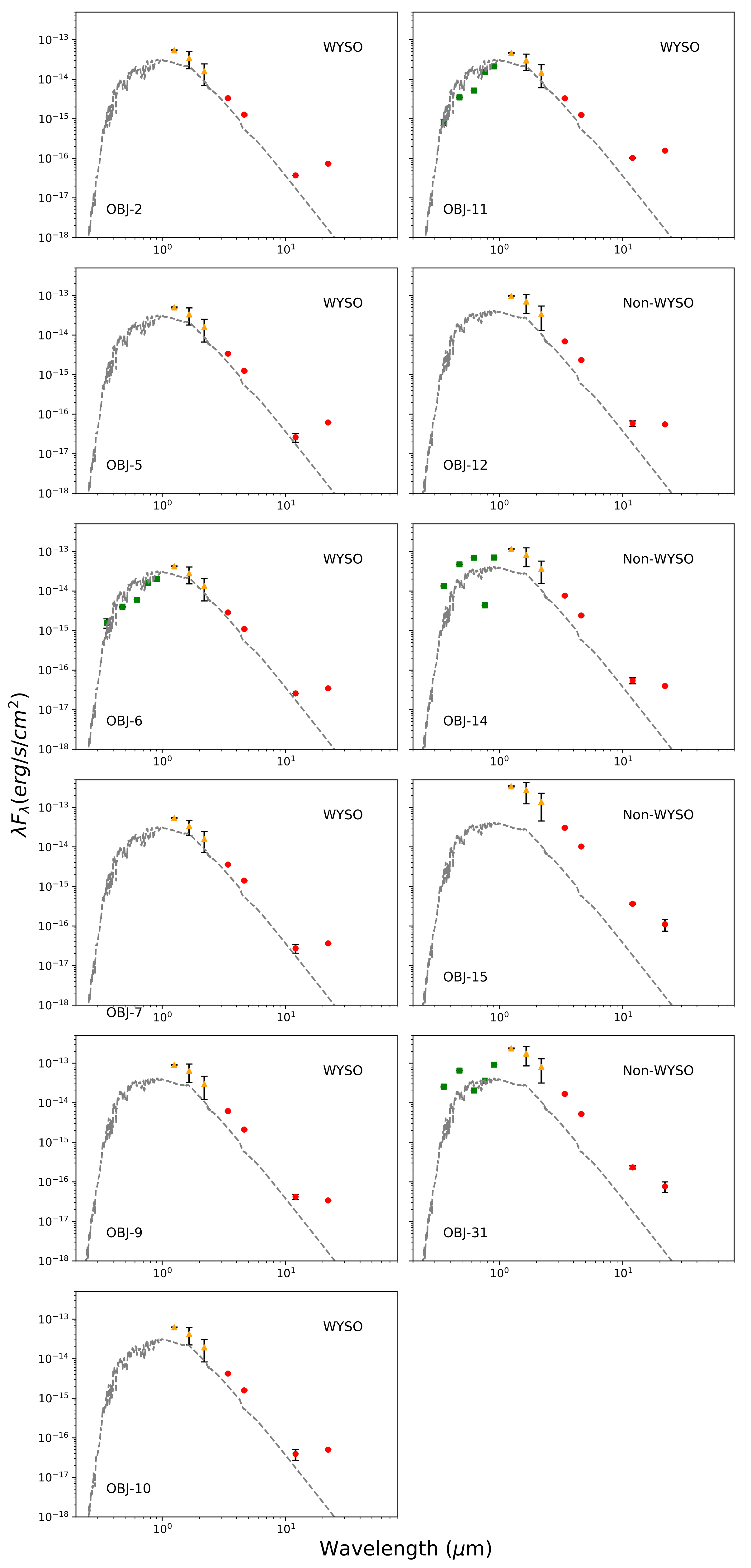}
    \caption{SEDs of the targets, showing the optical SDSS \textit{ugriz} (green squares; when available), 2MASS \textit{JHK} (yellow triangles), and WISE \textit{W1, W2, W3, W4} (red dots) fluxes. A Kurucz photosphere model for the given temperature is displayed (dashed lines) to clarify the possible presence of an IR excess.} %{\bf LaurentC4: Are there uncertainties/error bars on these measurements?}
    \label{fig:sed}
\end{figure}

\subsubsection{IR-excess} 

We also used photometric measurements from 2MASS \citep{cutri03} and WISE to identify IR excess that indicates the presence of a circumstellar disk and thus is a signature of youth. A condition used to classify the presence of IR excess is if their \textit{K} - \textit{W4} is larger than 0.2 \citep{wu13}. In Fig.~\ref{fig:ir_excess_wu}, we can see that all of the 11 Li-rich targets are well above the 0.2 threshold. 
%jct - can you plot all the 27 sources. Use some common, open symbol for these perhaps...

We also constructed the SEDs of each of the 11 targets using fluxes from WISE, 2MASS, and SDSS \citep{abdurrouf22} when available. Fig.~\ref{fig:sed} illustrates the SEDs and stellar photospheric models from \cite{Kurucz1992a} overlaid with each target. In all of the SEDs, it is clear that there is a notable increment in the flux at 2.4 $\mu$m, which indicates possible IR excess, even for the targets that are not classified as WISE YSO. 

\subsubsection{Li-poor targets}

Among the 16 observed targets which did not show strong Li or H$\alpha$ features, the target ID3018141830356350976 (OBJ-20) in particular, is located above the \citet{baraffe15} 1 Myr isochrone (see Fig.~\ref{fig:obj20cmd}), yet only has an EW(Li) of $81.93 \pm 25.25$ m\AA  (see Fig.~\ref{fig:obj20spectra}). Possible reasons why this source appears young in the CMD could be inaccurate reddening and extinction, and/or binarity. The other Li-poor targets, while showing other indications of youth, such as evidence of IR-excess, are generally located well below the Li-rich targets in CMDs, in closer proximity to 20 \& 30 Myr isochrones. They have likely either depleted most of their photospheric Li at this age, or else are older still but have inaccurate reddening and extinction, making them appear younger in the CMD. 

% Joe - you should also mention that the resulting EW(Ha), EW(Li) and RVs are given in Table 2.
%---

%This is particularly interesting because the label developed to identify YSOs turned out to contain numerous contaminants.

% Joe - rather than say "quite prominent" it would be better to quote the thresholds commonly used (EW(Ha) > 10 A, EW(Li) > 0.15 / 0.2 A) and cite papers that have also used them (like Armstrong et al. 2022)

% Joe - we should also estimate uncertainties for these EWs. In Armstrong et al. 2022 the way I did so was to take the RMS value of the EWs measured using the same procedure with the Gaussian profile of the CCF centred at five wavelengths either side of the Li line, which you can find more detail about in Jackson et al. 2018. Can you do the same in IRAF?

%The lithium EW for each of the 10 targets is notably large, ranging from 0.51 \AA\ for the smallest to 0.81 \AA\ for the largest. In contrast, the hydrogen-alpha (H$\alpha$) EW varies from a minimum of 4.56 \AA\ to a maximum of 18.61 \AA. These values suggest that the 10 targets are most likely YSOs, particularly due to the presence of lithium.
% Joe - this paragraph should be at the end of section 2

\subsection{Radial Velocities and 3D Traceback}
\label{section_traceback}
%\label{rvs}

The 11 targets with detected lithium were cross-correlated with matching synthetic spectra and RVs were determined from the position of the peak in the cross-correlation function (CCF) by fitting a Gaussian function. Synthetic spectra were produced using the MOOG spectral synthesis code \citep{Sneden2012a}, with \cite{Kurucz1992a} solar-metallicity model atmospheres, for $\log\ g$ = 4.0 from $T_{\rm eff}$ = 7000 K down to $T_{\rm eff}$ = 3500 K in 500 K steps.

To perform the RV measurement we use IRAF \texttt{rvsao} package \citep{mink98}. We computed the heliocentric velocity corrections using the IRAF \texttt{rvcorrect} task. RV uncertainties were determined empirically from the difference in RV between $n$ separate exposures of the same target ($\Delta v_r = (v_{r,{\rm max}}-v_{r, {\rm min}})/ \sqrt{n}$). Heliocentric RV ($v_r$) and uncertainties are given in Table \ref{results_table}.

\begin{table*}
\begin{center}
{\renewcommand{\arraystretch}{1.5}
\begin{tabular}{|p{1.0cm}|p{2.8cm}|p{1.2cm}|p{1.5cm}|p{1.6cm}|p{1.2cm}|p{1.4cm}|p{1.6cm}|p{0.7cm}|p{0.7cm}| }
\hline %31.175488, 13.418734, 12.840667, 15.261152, 15.067134, 25.157383, 14.126379, 16.427216, 11.816317, 13.69072
ID & Gaia ID & EW(Li) (m\AA) & EW(H$\alpha$) (m\AA) & $v_r$ \newline (km s$^{-1}$) & $v_{\rm 3D,ONC}$ (km s$^{-1}$) & $\tau_{\rm min,3D}$ (Myr) & $D_{\rm min,3D}$ (pc) & Score & Bona-fide \\
\hline
%OBJ-1 &4844159373956999936 &  \\
OBJ-2 & 3192134597649605376 & $750 \pm 82$ & $13260 \pm 99$ & $20.56 \pm 0.43$ & 31.17 & $4.40 \pm 0.27$ & $0.25 \pm 6.69$ & aI & \checkmark \\
%OBJ-3 &4817925576274600192 &  \\
%OBJ-4 &3212944607551376384 &  \\
OBJ-5 & 3207501131641282176 & $600 \pm 75$ & $15400 \pm 104$ & $15.87 \pm 0.42$ & 13.42 & $2.57 \pm 0.80$ & $19.47 \pm 7.01$ & aI &  \\
OBJ-6 & 3208291783581908608 & $616 \pm 68$ & $13700 \pm 105$ & $16.96 \pm 0.42$ & 12.84 & $4.34 \pm 1.07$ & $2.32 \pm 9.09$ & bI & \checkmark \\
OBJ-7 & 3207022053810350976 & $529 \pm 87$ & $9310 \pm 126$ & $14.96 \pm 1.02$ & 15.26 & $3.19 \pm 0.72$ & $8.01 \pm 6.81$ & aI & \checkmark \\
%OBJ-8 &2957497235735128448 &  \\
OBJ-9 & 2984454031031531008 & $643 \pm 42$ & $6220 \pm 57$ & $19.78 \pm 1.01$ & 15.06 & $4.68 \pm 0.31$ & $5.02 \pm 3.11$ & aI* & \checkmark \\
OBJ-10 & 3216889827071056896 & $440 \pm 102$ & $5425 \pm 127$& $31.44 \pm 2.73$ & 13.38 & $1.52 \pm 0.92$ & $1.67 \pm 10.57$ & bI & \checkmark \\
OBJ-11 & 3215804677813294976 & $513 \pm 62$ & $10510 \pm 86$ & $3.54 \pm 0.81$ & 25.16 & $1.08 \pm 0.71$ & $4.28 \pm 18.59$ & bI & \checkmark \\
OBJ-12 & 3012142379518284288 & $810 \pm 62$ & $4560 \pm 77$ & $21.88 \pm 2.42$ & 14.13 & $2.22 \pm 0.46$ & $4.28 \pm 7.09$ & bI* & \checkmark \\
%OBJ-13 & 3015308629408804608 & \\
OBJ-14 & 3219378365481960832 & $696 \pm 63$ & $4610 \pm 85$ & $38.78 \pm 4.41$ & 16.43 & $2.11 \pm 0.63$ & $25.63 \pm 5.26$ & bI! &  \\
OBJ-15 & 3216174629116142336 & $520 \pm 46$ & $186100 \pm 88$ & $27.18 \pm 1.67$ & 11.82 & $1.85 \pm 0.34$ & $10.54 \pm 4.66$ & bI &  \\
%OBJ-16 &2888109908763598976 &  \\
%OBJ-17 &3011187006993509504 &  \\
%OBJ-18 &3316420643274767488 &  \\
%OBJ-19 &3315632671394273024 &  \\
%OBJ-20 &3018141830356350976 &  \\
%OBJ-21 &2932903703242234112 &  \\
OBJ-22 & 3017382033474172800 & $549 \pm 48$ & $8710 \pm 60$ & $14.57 \pm 1.28$ & 13.69 & $0.97 \pm 0.30$ & $0.52 \pm 3.64$ & bI & \checkmark \\
%OBJ-23 &3023944262453551232 &  \\
%OBJ-24 &2915784994393795456 &  \\
%OBJ-25 &2885209740687428224 &  \\
%OBJ-26 &3317517165606496256 &  \\
%OBJ-27 &5709959085012172928 &  \\

\hline
\end{tabular}}
\end{center}
\setlength{\belowcaptionskip}{-10pt}
\setlength{\textfloatsep}{0pt}
\caption{Results of spectral and traceback analysis for targets with spectroscopic youth indicators. Columns are; Gaia DR3 unique ID number, equivalent width of Li, equivalent width of H$\alpha$, heliocentric radial velocity, time since closest approach to the ONC, distance of closest approach to the ONC, \protect\citet{farias20} score for youth criteria met and flag for traceback consistent with originating from the ONC. }
\label{results_table}
\end{table*}
% Joe - I think this table should also include the results for the targets that did not meet the youth criteria, 
%Calculate RV for others - ?

%\subsection{Traceback Analysis}
%\label{section_traceback}

Now that we have confirmed 11 of our runaway candidates as YSOs via spectroscopic youth signatures and have measured $v_r$ for them, we can trace back their past trajectories in 3D to confirm their possible origin in the ONC and estimate the time since their ejection from the cluster. 

%\subsubsection{Traceback Calculation}

We begin, as in \citet{farias20}, by defining the reference frame of the ONC. We adopt the central coordinates of the cluster as $RA = 05h35m16.26s, Dec = -05d23m16.4$s and the distance as 403 pc \citep{dario16}. We adopt an ONC proper motion of $\mu_{\alpha*} =  1.43 \pm 0.14$ mas yr$^{-1}$ and $\mu_{\delta} = 0.52 \pm 0.12$ mas yr$^{-1}$ from \cite{kuhn19} and cluster mean $v_r$ of $26.4 \pm 1.6$ km s$^{-1}$ from \citet{farias20}, \cite{dario14}, and \cite{hoogerwerf01}.
% Joe - can you find the original references for these values? The RV is probably from Da Rio et al. 2014
%--
For the calculation of 3D trajectories, observed astrometry and radial velocities for each runaway candidate, as well as the ONC frame, were transformed into 3D Cartesian positions and velocities $X, Y, Z, U, V, W$ (along with their associated uncertainties) to eliminate the need for corrections of perspective effects caused by a spherical coordinate system and the motion of the Sun therein. 

As in \citet{farias20}, 3D traceback is performed using vector algebra. To find the time and distance of closest approach to the center of the ONC we use
\begin{equation}
    \tau_{\rm min,3D} = -\frac{(X_*-X_0)\dot(V_*-V_0)}{|V_*-V_0|^2}
\end{equation}
and
\begin{equation}
    D_{\rm min,3D} = |(X_*-\tau_{\rm min,3D}V_*)-(X_0-\tau_{\rm min,3D}V_0)|,
\end{equation}
where $X_*$ and $V_*$ are the 3D position and velocity of the star and $X_0$ and $V_0$ are the 3D position and velocity of the ONC.

We find a range of 3D ejection timescales among these 11 YSOs, ranging from $0.97 - 4.68\:$Myr with a typical precision of $\sim0.5\:$Myr, as well as a range of closest approach distances from $0.25 - 25.6\:$pc with a typical precision of $\sim6\:$pc. 
Table~\ref{results_table} summarizes these results, including measured equivalent widths, radial velocities, and 3D traceback properties. 

\begin{figure*}
    \centering
    \includegraphics[width=1.0\textwidth]{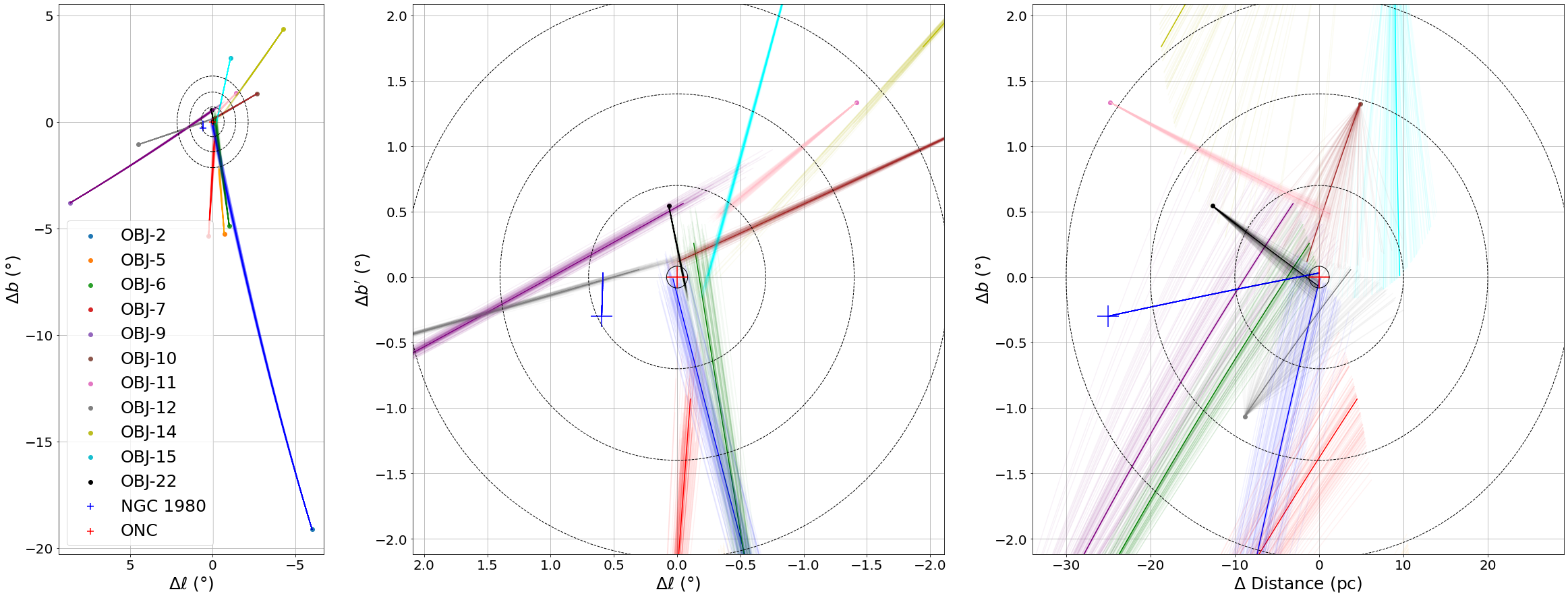}
    \caption{\textit{Left:} Positions relative to the ONC of spectroscopically confirmed YSOs and their plane-of-sky trajectories from the position of 3D closest approach to the ONC. Solid lines indicate the median trajectory, faint lines indicate 100 trajectories per runaway calculated with random contributions from their proper motion and radial velocity errors, assuming a fixed distance. Dashed circles indicate concentric radii centred on the ONC at radial distances of 10, 20 \& 30 pc. The blue cross indicates the central position of NGC 1980 and its median trajectory relative to the ONC is also plotted. \textit{Middle:} Zoomed-in on the region around the ONC. \textit{Right:} Line-of-sight distance (pc) against Galactic latitude.}
    \label{fig:trajectory}
\end{figure*}

In Fig.~\ref{fig:trajectory} we plot the positions of the spectroscopically confirmed YSOs relative to the ONC ($\Delta\ell, \Delta b, \Delta Distance$), with solid lines to indicate their mean trajectory relative to the ONC from their point of closest approach, and faint lines to indicate the uncertainty on their trajectories, produced by Monte Carlo with 100 iterations each time adding perturbations randomly sampled from their proper motion and RV errors. The dashed circles indicate 10, 20 and 30 pc radii centered on the ONC. We also indicate the position and past motion of NGC 1980 relative to the ONC.

From the above results, we see that most of the 11 sources satisfy 3D traceback, i.e., with $D_{\rm min,3D}$ consistent with zero within $3\sigma$, with the exception of OBJ-14. We note the radius of the ONC is estimated to be about $2.5 - 3$ pc \citep{dario14,kroupa18}, therefore we can discard runaway candidates who do not trace back to a closest approach distance within this radius within their uncertainties. From this filtering process, we conclude that OBJ-14 is least likely to have been ejected from the ONC. In addition, OBJ-5 and OBJ-15 have minimum ONC approach distances that are about $2-3\sigma$ deviant from zero, which raises doubts about their origin in the ONC. Thus our finalized, highest confidence sample of ONC runaways consists of 8 sources: OBJ-2, 6, 7, 9, 10, 11, 12, 22, as indicated by the last column of Table.~\ref{results_table}.

% Joe - even though you give the results in Table 2. you should still summarise them in a paragraph here, quoting the mean value and mean uncertainty of each

% Joe this is where we should have the plane-of-sky traceback plot with trajectories for each runaway (so that the trajectory ends at the closest approach distance in 2D) and with shaded regions for uncertainties. Perhaps we could have two panels, the first showing the whole area of the sky where the runaways are located and their entire trajectories, the second zoomed into the ONC so that we can see each runaway's 2D point of closest approach in comparison to the concentric radii around the ONC center as dashed lines.

\section{Discussion}
\label{section_discussion}

\begin{figure*}
    \centering
    \includegraphics[width=\textwidth,center]{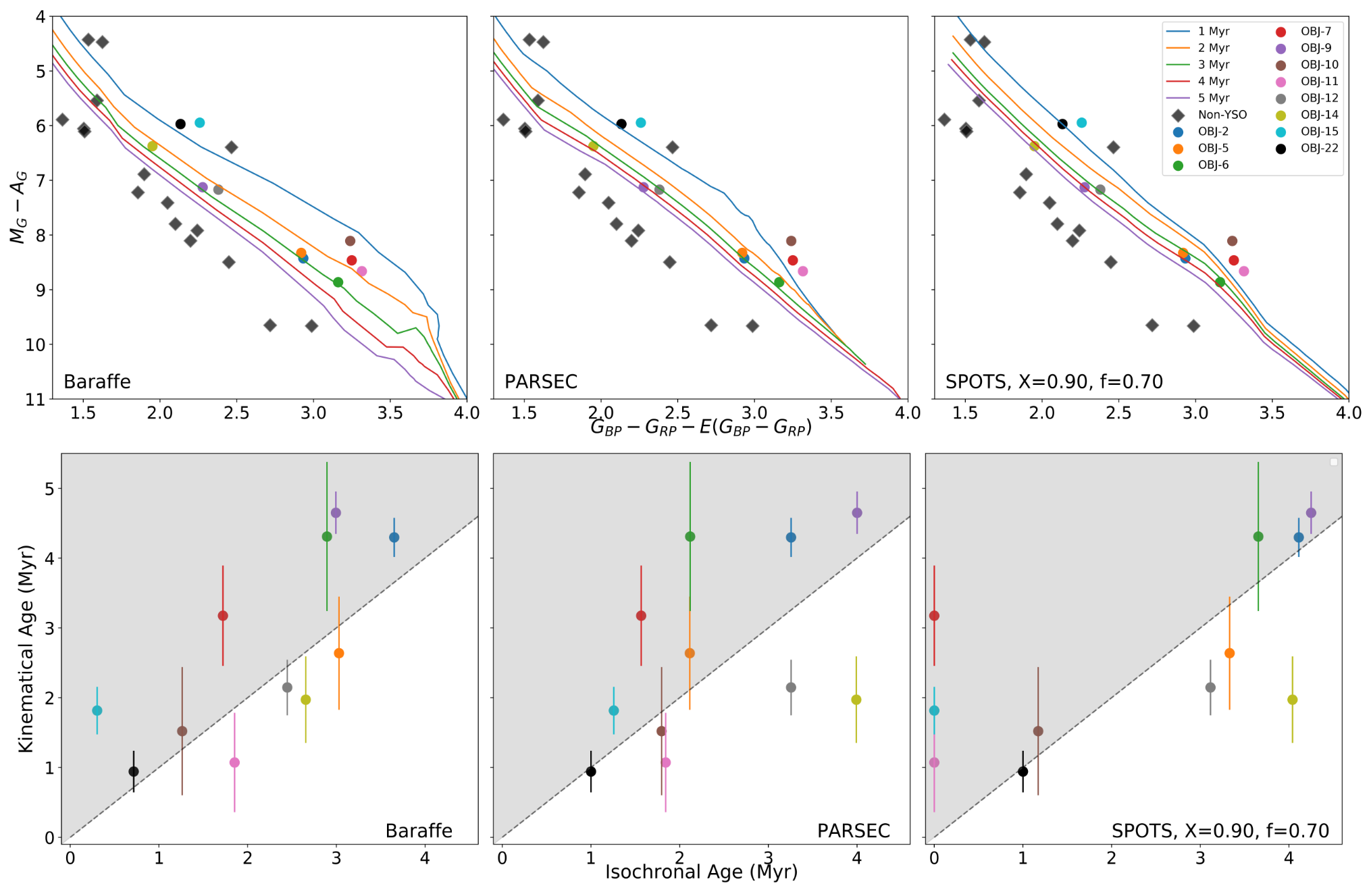}
    \caption{\textit{Top:} Gaia DR3 BP-RP colour - MG absolute magnitude diagram for spectroscopically confirmed YSOs in our sample. Extinction AG and reddening E(BP-RP) are estimated per source from Gaia. Overlaid are \protect\citet{baraffe15}, PARSEC \protect\citep{marigo17} and SPOTS \protect\citep[][x=0.9, f=0.7]{somers20} isochrones for 1, 2, 3, 4 and 5 Myr. \textit{Bottom:} Plot of isochronal ages estimated by CMD position relative to \protect\citet{baraffe15}, PARSEC \protect\citep{marigo17} and SPOTS \protect\citep[][x=0.9, f=0.7]{somers20} isochrones - ejection timescales for spectroscopically confirmed YSOs.  } %Baraffe (\textit{top}), PARSEC (\textit{middle}), and SPOTS (\textit{bottom}) isochrones for 1, 2, 3, 4 and 5 Myr.}
    \label{fig:cmds}
\end{figure*}

\subsection{Comparison of Kinematic Ejection Ages with Isochronal Ages}

%We have gathered several pieces of evidence indicating that the 10 targets are most likely YSOs. The availability of optical spectra allowed us to also measure radial velocities, enabling us to assess 3D traceback of the targets to determine if they were ejected from the ONC, making them candidates of genuine ONC runaways. We calculated the closest distance that the candidates had to the ONC in the past. For most of our spectroscopically confirmed YSOs, their 3D traceback is consistent with an origin in the ONC, meaning they were much closer to the ONC in the past.

% Joe - this summary paragraph sounds like it belongs in the Summary and Conclusions section. Make sure that you are not repeating the same text in different sections

%We noted that there are objects with a closest approach well below 10 pc, around 10 pc, and above 10 pc. At this point, we consider those well below 10 pc to be "likely", around 10 pc to be "possible", and above 10 pc to be "unlikely" candidates for having been ejected from the ONC. The "likely" candidates also happen to be around or below 5 pc, and based on \cite{kroupa18}, nominal radius of the ONC is 2.5 pc, it strengthens our guess of them being likely ejected from the ONC. From this filtering process, we identified three objects that are unlikely to have been ejected from the ONC: OBJ-2, OBJ-5, and OBJ-14. Thus, we are left with seven likely and possible candidates to have been ejected from the ONC. 

%are unlikely to have been ejected from the ONC. %Thus, we are left with six likely and possible candidates to have been ejected from the ONC. 

True YSO runaways cannot have an ejection timescale longer than the age of the YSO itself. To double-check the feasibility of these ejection timescales, we estimated the isochronal ages of our candidates using Baraffe \citep{baraffe15}, PARSEC \citep{marigo17}, and SPOTS (with X=0.90 and f=0.70) \citep{somers20} models (see Fig.~\ref{fig:cmds}). We took into account extinction and reddening for each source, taken from \cite{farias20}. We note that isochronal ages for low-mass YSOs estimated in this way can vary significantly depending on the stellar evolution models used (Table~\ref{isochrone_table}). For a given position on the Color-Magnitude Diagram (CMD), models that include magnetic activity, e.g., SPOTS, tend to predict older ages than models without such effects. However, at very low masses and at very young ages, the situation can reverse, with the Baraffe models predicting older ages.

Given the possibility of large systematic errors in the isochronal age estimates, we show the results from use of each model, treating this an approximate way of estimating a range of possible ages. However, the formal uncertainties in the isochronal ages are difficult to assess.

%we treat this variation as an age range for each candidate, giving us a range of age estimates rather than a single age estimation. 
% Joe - more detail is needed to explain where the extinction and reddening values come from, are they from Farias et al. 2020?
%More detail is needed to explain the isochronal age interpolation. Also we should reference more papers concerning the discrepancies between different models, particularly for young low-mass stars, such as Cao et al. 2022

Nevertheless, we then compared our ejection timescales to the isochronal age estimates (Fig.~\ref{fig:cmds}). YSOs with isochronal ages similar or older than their ejection timescales are more likely to be old enough to have been ejected from the ONC, given their current position and velocity. On the other hand, YSOs with isochronal ages younger than their ejection timescales should be excluded from having been ejected from the ONC. We have shaded this “forbidden zone” in grey in Fig.~\ref{fig:cmds}. However, we note that uncertainties in the ejection and isochronal ages could cause stars to scatter into this forbidden zone.  %Inversely, YSOs with isochronal ages significantly younger than their ejection timescales are less likely to have been ejected from the ONC, as indicated by the shaded regions in Fig.~\ref{fig:cmds}.
%Thus, they are very likely to be actual runaway stars from the ONC. 
From this assessment, OBJ-7 is the most suspect, since its oldest isochronal age estimate (from Baraffe) is only about 60\% of its ejection age. Still, even here, given potential systematic uncertainties in isochronal ages, we consider that this source could still be an ONC runaway. For the remaining sources, we find that isochronal age estimates are generally consistent with their ejection ages.

%our candidates are further reduced because OBJ-6 and OBJ-7 specifically have isochronal ages lower than their ejection timescales, making them less likely to have been ejected from the ONC. However, in the case of OBJ-6, there is still overlap between the isochronal age and the ejection age. We decided to further discard OBJ-7 but still keeping OBJ-6 in our final candidates. Finally, we have six new likely candidates to be runaways from the ONC, based on the closest approach and age estimates: OBJ-2, OBJ-6, OBJ-9, OBJ-11, OBJ-12, and OBJ-22.

% Joe - there should be mention of the possibility that candidate runaways with ejection timescales larger than their isochronal ages could have originated outside the ONC, from another cluster in the vicinity, or from an extended population of young stars (like an OB association). We could have a look for other young clusters and O/B stars in the vicinity of the ONC (using catalogues) that could possibly be the origin of such candidates. If we find no other suitable origin (no other cluster that intersects with a candidates past trajectory), then we can safely assume that the isochronal ages are underestimated and that the candidates are confirmed ONC runaways.

\begin{table*}
\begin{center}
{\renewcommand{\arraystretch}{1.5}
\begin{tabular}{|p{1.2cm}|p{3.5cm}|p{0.9cm}|p{1.1cm}|p{1.0cm}|p{0.9cm}|p{1.1cm}|p{1.0cm}|}
\hline
ID & Gaia ID & Baraffe age (Myr) & Baraffe mass (M$_\odot$) & PARSEC age (Myr) & PARSEC mass (M$_\odot$) & SPOTS age (Myr) & SPOTS mass (M$_\odot$) \\
\hline
%4844159373956999936 &  \\
OBJ-2 & 3192134597649605376 & 3.65 & 0.24 & 3.25 & 0.28 & 4.11 & 0.28\\
%4817925576274600192 &  \\
%3212944607551376384 &  \\
OBJ-5 & 3207501131641282176 & 3.03 & 0.24 & 2.11 & 0.28 & 3.33 & 0.28\\
OBJ-6 & 3208291783581908608 & 2.89 & 0.18 & 2.12 & 0.19 & 3.65 & 0.22\\
OBJ-7 & 3207022053810350976 & 1.72 & 0.17 & 1.56 & 0.35 & $\ll$1 & 0.20\\
%2957497235735128448 &  \\
OBJ-9 & 2984454031031531008 & 2.99 & 0.42 & 4.00 & 0.58 & 4.25 & 0.51\\
OBJ-10 & 3216889827071056896 & 1.26 & 0.18 & 1.79 & 0.63 & 1.17 & 0.43 \\
OBJ-11 & 3215804677813294976 & 1.85 & 0.16 & 1.84 & 0.36 & $\ll$1 & 0.19\\
OBJ-12 & 3012142379518284288 & 2.45 & 0.39 & 3.25 & 0.51 & 3.11 & 0.46\\
%3015308629408804608 &  \\
OBJ-14 & 3219378365481960832 & 2.65 & 0.56 & 3.99 & 0.69 & 4.03 & 0.75\\
OBJ-15 & 3216174629116142336 & 0.30 & 0.50 & 1.25 & 0.73 & $\ll$1 & 0.45\\
%2888109908763598976 &  \\
%3011187006993509504 &  \\
%3316420643274767488 &  \\
%3315632671394273024 &  \\
%3018141830356350976 &  \\
%2932903703242234112 &  \\
OBJ-22 & 3017382033474172800 & 0.71 & 0.45 & 1.00 & 0.55 & 1.00 & 0.53\\
%3023944262453551232 &  \\
%2915784994393795456 &  \\
%2885209740687428224 &  \\
%3317517165606496256 &  \\
%5709959085012172928 &  \\

\hline
\end{tabular}}
\end{center}
\setlength{\belowcaptionskip}{-10pt}
\setlength{\textfloatsep}{0pt}
\captionsetup{justification=centering}
\caption{Isochronal ages and masses of each spectroscopically confirmed YSO estimated from \protect\citet{baraffe15}, PARSEC \protect\citep{marigo17} and SPOTS \protect\citep[][x=0.9, f=0.7]{somers20} isochrones.}
\label{isochrone_table}
\end{table*}

\subsection{Confirmed ONC Member}

One of our runaway candidates, OBJ-22 (Gaia ID 3017382033474172800), also known as V* V1781 Ori,  has been reported by \citet{rebull01} as an M2.5 star member of the ONC. The interpolated temperature suggests that this object has a temperature of approximately 3700 K. The age estimation for this object ranges from 0.4 to 1.00 Myr. Based on our traceback calculation, the star should have been ejected $0.97 \pm 0.30$ Myr ago, with the closest approach to the ONC of $0.52 \pm 3.64$ pc, which is the second smallest among our candidates. Considering both the closest approach and timescales of the object, it is evident that OBJ-22, or V*1781 Ori, is not only a member of the ONC but also a relatively recent runaway star. 
% Joe - since this is a much more recently ejected runaway than the oldest we can comment on the spread of ejection timescales and compare to predictions from N-body simulations that track ejections, such as Farias et al. 2018

% Joe - once you have added the 3D velocities relative to the ONC for these candidates we can check that this star has a velocity consistent with a real dynamical ejection rather than just being a low-velocity unbound YSO

\begin{figure*}
    \centering
    \includegraphics[width=1.0\textwidth]{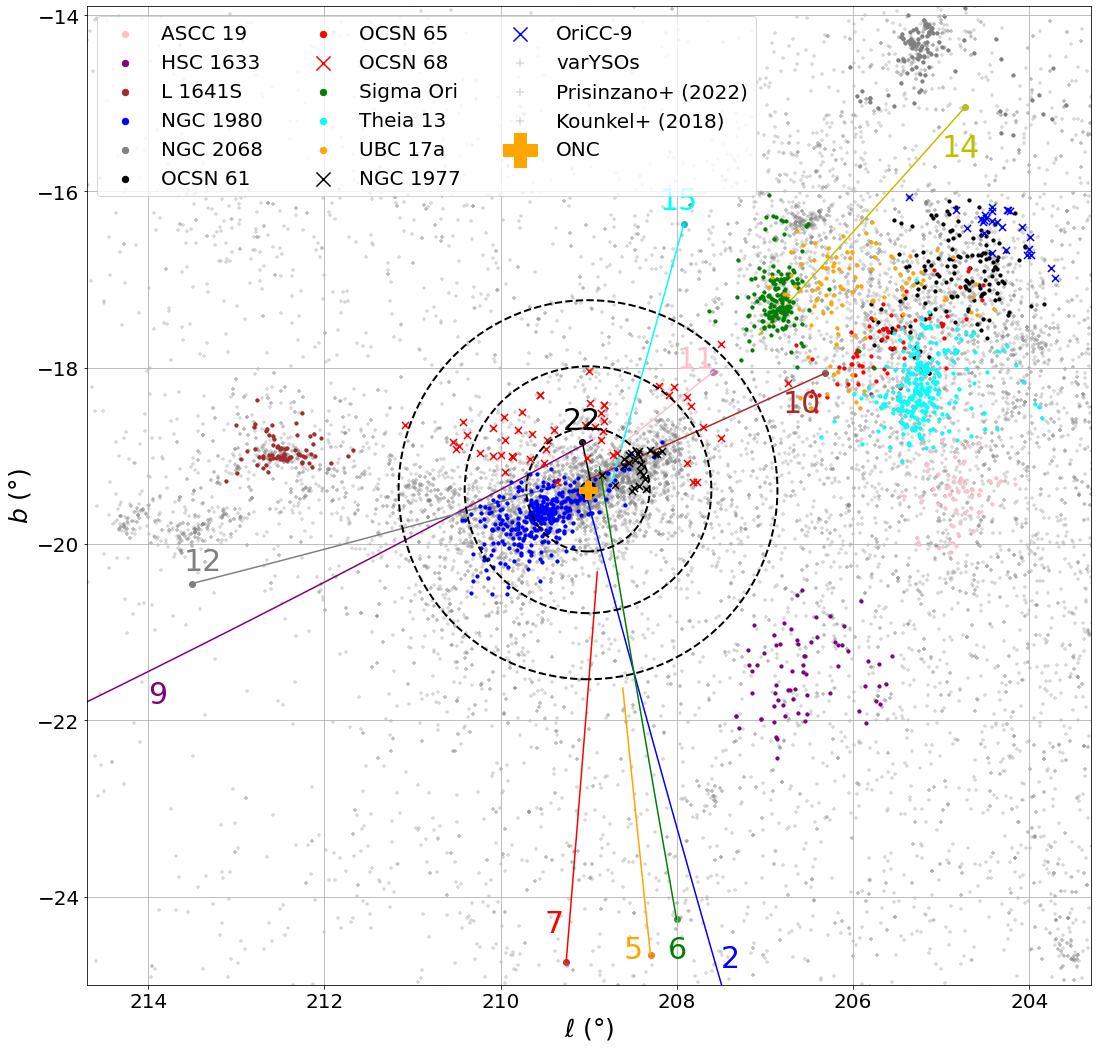}
    \caption{Positions in Galactic coordinates of spectroscopically confirmed YSOs and their plane-of-sky trajectories from the position of closest approach to the ONC. Dashed circles indicate concentric radii centred on the ONC at radial distances of 10, 20 \& 30 pc. Coloured points and crosses indicate members of other nearby young ($<15$ Myr) clusters in the region at mean distances between 350 - 420 pc from \protect\citet{hunt23} and \protect\citet{kounkel22}, grey points are YSOs from the Gaia DR3 variable source catalogue \protect\citep{marton23}, APOGEE targets from \protect\citet{kounkel18} or candidate young stars identified by \protect\citet{prisinzano22}.}
    \label{fig:clusters}
\end{figure*}

\subsection{Alternative Origin Clusters}
\label{subsection_clusters}

Apart from the ONC, the Orion star-forming region is a large complex consisting of many sub-regions ranging from sparse associations to dense clusters. In order to be {\bf more} certain that the ONC is the origin of our runaways candidates we need to consider whether their trajectories intersect with any other young clusters, especially for candidates where their ONC ejection timescale is greater than their isochronal age estimate.

We check the recent cluster catalogue of \citet{hunt23} for other young ($<$15 Myr) clusters in the region, limiting to clusters with a mean distance in a similar range as that of our YSO runaway candidates ($\sim$ 350 - 420 pc). The distribution of their members on the sky is shown in Fig.~\ref{fig:clusters} along with the positions and trajectories of the YSOs and the position of the ONC. We note that none of our YSO runaway candidates are included as members of any of these clusters in this catalogue. We also plot other candidate YSOs in the region, either from the Gaia DR3 variable source catalogue \protect\citep{marton23}, APOGEE targets from \protect\citet{kounkel18} or candidate young stars identified by \protect\citet{prisinzano22}, as grey points.

Using cluster central positions, proper motions, distances and radial velocities from the catalogue, we repeat the 3D traceback analysis for our YSOs relative to these clusters. In particular, we find that OBJ-14 has a $D_{\rm min,3D} = 8.85 \pm 5.31$ pc and $\tau_{\rm min,3D} = 1.24 \pm 0.45$ Myr relative to the $\sigma$ Ori cluster (green in Fig.~\ref{fig:clusters}), making it much more likely to be a runaway from the $\sigma$ Ori cluster \citep[2.8 Myr, 46 members; ][]{kounkel22} than the ONC. Also, we find that OBJ-9 has a $D_{\rm min,3D} = 3.05 \pm 6.04$ pc and $\tau_{\rm min,3D} = 4.26 \pm 0.32$ Myr relative to the NGC 1980 cluster (blue cross in Fig.~\ref{fig:trajectory}, blue in Fig.~\ref{fig:clusters}), though both of these values are similar to those calculated when tracing from the ONC. We also find that OBJ-6 has a $D_{\rm min,3D} = 5.96 \pm 12.39$ pc and $\tau_{\rm min,3D} = 3.79 \pm 0.14$ Myr relative to NGC 1980 \citep[3.1 Myr, 34 members; ][]{kounkel22}. The argument can be made that the ONC, being the more massive and dense cluster \citep[$>10^4 M_\odot {\rm pc}^{-3}$; see ][]{hillenbrand97}, will have many more dynamical interactions between its members and thus will eject many more runaways than NGC 1980, making OBJ-9 \& 6 more likely to have originated from the ONC. 

To better quantify this hypothesis, we make an estimate of the central density of the NGC 1980 cluster. We have taken the sample of 364 members of NGC 1980 from the \citet{hunt23} catalogue and estimated masses for these by comparison to a $5Myr$ PARSEC \citep{marigo17} isochrone, which we then use to extrapolate the total cluster mass ($\sim340 \pm 40 M_\odot$) and number of stars ($\sim1050$) from a \citet{maschberger13} IMF. We then use the total number of stars to scale the radial profile of NGC 1980 members from \citet{hunt23}. We estimate a central projected number density of $\sim10^2 {\rm pc}^{-2}$ in NGC 1980, which is $\sim5 - 10$ times smaller than typical estimates for the ONC (see Fig. 8 of \citealt{farias23}; $5\times10^2 – 10^3 {\rm pc}^{-2}$).
%, also \citealt{hillenbrand97}; $10^4 pc^{-3}$).
Assuming, conservatively, that the difference in 3D number density is at a similar level (note, that estimates of the 3D number density in the center of the ONC are at a level of $\sim 10^4\:{\rm pc}^{-3}$; \citealt{hillenbrand97}),
then this difference would then imply $25 – 100$ times lower rate of dynamical interactions in NGC 1980 than in the ONC. Furthermore, considering the total numbers of stars ($\sim1050$ for NGC 1980, $\sim3500$ for the ONC; \citealt{hillenbrand97}), implies $\sim80 – 350$ times fewer ejections from NGC 1980 compared to from the ONC. Thus, any candidate runaway whose trajectory is consistent with originating in either NGC 1980 or the ONC is much more likely to have originated from the ONC.
%on the basis of total number of runaways produced by each cluster alone.  }

We also check for matches between our YSO runaway candidates and the runaway candidate catalogue of \citet{kounkel22} and the clusters/subclusters they trace from in the plane-of-sky. OBJ-2 is included as a possible runaway from their Rigel subcluster \citep[5.7 Myr, 24 members; ][]{kounkel22}, OBJ-10 is included as a possible runaway from epsilon Ori-2 \citep[3.8 Myr, 33 members; ][]{kounkel22}, OBJ-11 is included as a possible runaway from NGC 1977 \citep[2.6 Myr, 22 members; ][]{kounkel22}, OBJ-12 from LDN 1647 \citep[2.1 Myr, 86 members; ][]{kounkel22} and OBJ-14 from OriCC-9 \citep[3.6 Myr, 26 members; ][]{kounkel22}. None of our YSO runaway candidates match with their cluster/subcluster members. 

LDN 1647 has a mean parallax of 2.284 mas ($\sim$ 437 pc), and has 23 members with RVs available from SOS \citep{tsantaki22} giving a mean cluster $v_r$ of 20.55 km s$^{-1}$ with a dispersion of 1.01 km s$^{-1}$. OBJ-12 has a distance of $386.4^{+4.5}_{-4.3}$ pc \citep{bailerjones21} and an $v_r$ of $21.88 \pm 2.42$ km s$^{-1}$ and is thus unlikely to have originated from LDN 1647 as a $\sim$3 Myr old YSO would need to travel at $\sim$17 km s$^{-1}$ relative to the cluster to achieve a relative distance of $\sim$50 pc in the line-of-sight. In fact, OBJ-12 is likely to be moving toward LDN 1647 in the line-of-sight and therefore cannot have originated from it. %{\bf LaurentCX: I would add units pc for the distance and km per second for the RV} 

NGC 1977 has a mean parallax of 2.572 mas ($\sim$ 389 pc), and has 8 members with RVs available from SOS giving a mean cluster $v_r$ of 30.10 km s$^{-1}$ with a dispersion of 0.91 km s$^{-1}$. OBJ-11 has a distance of $370.9^{+10.8}_{-10.3}$ pc \citep{bailerjones21} and an $v_r$ of $3.54 \pm 0.81$ km s$^{-1}$ and is thus moving away from the cluster at a relative velocity of 26.56 km s$^{-1}$. Therefore, if OBJ-11 was ejected from NGC 1977 it would have an ejection timescale of $\sim0.68$ Myr.

OriCC-9 and epsilon Ori-2 have only 2 members with RVs from SOS \citep{tsantaki22} and the Rigel subcluster only 1, making it difficult to determine their group 3D kinematics. Thus, we cannot entirely dismiss these groups as possible origins of OBJ-2, OBJ-10 and OBJ-14. But again, we reiterate that as the most massive cluster, the ONC is the more likely origin for candidate runaways in general when we cannot compare precise 3D trajectories.

We also note that OBJ-5, OBJ-6, OBJ-7 and OBJ-22 are included among the 14832 members of the Orion cluster 606 of \citet{prisinzano22}, while our other YSO runaway candidates are not included in any cluster of theirs. However, as shown in Fig.~\ref{fig:clusters}, OBJ-5, OBJ-6, OBJ-7 are not clearly affiliated at present with any dense cluster or substructure within the Orion region. Thus, given their 3D trajectories, and relative velocities ($v_{\rm 3D,ONC} >$ 12 km s$^{-1}$), we can conclude that they are likely ONC runaways rather than being members of the sparsely distributed and dispersing population.

Apart from the likely origin of OBJ-14 from the $\sigma$ Ori cluster, we do not find more likely origin clusters in the Orion region for our runaway candidates than the ONC, on the basis of 3D closest approach analysis and given the much greater rate of ejections expected from the ONC compared to other clusters in the region. In particular, we do not find {\bf more} likely alternative clusters of origin for our confirmed YSO candidates with the longest ejection timescales OBJ-2, OBJ-6 \& OBJ-9, strengthening the evidence that they originated from the ONC.

\subsection{New Oldest Runaways and Implications for Star Formation Efficiency per Free-fall Time}
\label{subsection_isochrones}

Among our very likely runaway candidates, we noted that several %objects 
%not only have a well-constrained closest approach distance to the ONC, but also possess 
have relatively old ejection ages.
%In particular, OBJ-6 
%has an ejection age of $4.34 \pm 1.07\:$Myr
%(and isochronal age estimated to be 3.65 Myr from comparison to SPOTS models),
%while OBJ-9 
%has an ejection age of $4.68 \pm 0.31\:$Myr (and isochronal age of 4.25 Myr). 
In particular: OBJ-2 has an ejection age of $4.40 \pm 0.27\:$Myr (and SPOTS isochronal age of 4.11 Myr);
OBJ-6 
%a candidate with the second-lowest closest approach of 2.04 pc, 
has an ejection age of $4.34 \pm 1.07\:$Myr
%However, its 
(and SPOTS isochronal age of 3.65 Myr);
and OBJ-9 
%with a closest approach just after OBJ-6 in our candidates at 4.29 pc, 
has an ejection age of $4.68 \pm 0.31\:$Myr (and SPOTS isochronal age of 4.25 Myr). 
%OBJ-12, with a closest approach of 5.98 pc, has an ejection age lower than its isochronal age, $2.22 \pm 0.46$ Myr and 3.25 pc, respectively.
%OBJ-6 and OBJ-9 are on the critical boundary for our runaway age criteria (ejection age vs. isochronal age), while OBJ-12 is well within the criteria. However, OBJ-6 has the closest approach among our candidates, making it still worth considering as a potential actual runaway from the ONC. 
%The same applies to OBJ-9, whose closest approach is just after OBJ-6. 
%Based on these objects, we found that their 
These ejection ages are longer than those of the oldest known ONC runaways to date, i.e., $\mu$ Col and AE Aur, with an ejection timescale of 2.5 Myr \citep{hoogerwerf01}.

%, up to almost twice as previously known (2.5 Myr). Although the ejection and isochronal age may still possess uncertainty, each mentioned candidate for the new oldest runaways agrees in terms of isochronal age to be larger than the currently known age of the ONC. 
%Therefore, in any case, from this work, we can suggest possible new lower limits for the ONC age. 

Thus, assuming that it indeed formed in the ONC and was then ejected, OBJ-9 would set a new record for the oldest detected runaway star from the ONC. It also establishes a new lower limit for the age of the ONC itself, i.e., $4.68 \pm 0.31\:$Myr.

This new lower limit for the age of the ONC has implications for the global star formation history, including star formation rates and efficiencies from the natal gas clump, which are important constraints on models of star cluster formation.
%the timescales that govern the formation of the cluster itself, thus affecting the constraints of the cluster formation. 
Following \citet{tan06} and \citet{dario14}, we estimate the star formation efficiency per free fall time as $\epsilon_{\rm ff} = 0.9 \epsilon_* t_{\rm ff}/t_{\rm form,90}$. We adopt $t_{\rm ff}\simeq 0.5\:$Myr, based on the dynamical mass model of \citet{dario14} at about the half-mass radius of 1.3~pc. Similarly, based on their models, we assume $\epsilon_*$, i.e., the overall fraction of gas that has formed stars, to be $\simeq 0.5$ (including allowance for some already expelled gas). We set the timescale for cluster formation, i.e., to form 90\% of the stars, to be our longest ejection age, i.e., 4.68~Myr. Thus we estimate $\epsilon_{\rm ff} \simeq 0.048$, which is valid at the half-mass radius scale. The value of $\epsilon_{\rm ff}$ is expected to be smaller at interior radii, where the densities are higher and the free-fall time decreases. 

We note that our estimate of $\epsilon_{\rm ff}$ is similar to that of \citet{dario14}, which is a reflection of the fact that our oldest ejection age is similar to their estimate of isochronal age spreads in the ONC. It should also be noted that the ejection of OBJ-9 4.7~Myr ago would already have required the presence of a dense, relatively massive stellar system, i.e., at least a triple system from which it is typically the lowest mass member, i.e., OBJ-9, that is ejected. Such a triple system would have required some time to form, i.e., if it involves $10\:M_\odot$ of stars forming from a core in a $\Sigma_{\rm cl}=0.3\:{\rm g\:cm}^{-2}$ environment, then a formation time of $2.4\times 10^5\:$yr is expected \citep{mckee03}. In addition, from statistical considerations it is likely that additional stars in the proto-ONC would have already been forming before the particular system that ejected OBJ-9. Thus star formation is likely to have been proceeding in the ONC for longer than 4.7~Myr. Since star formation continues today in the ONC and is expected to do so at least into the near future, it is reasonable to estimate a total duration of star cluster formation that is $>5\:$Myr for the ONC. Our above estimate for $\epsilon_{\rm ff} \simeq 0.048$ already accounts for a fraction of star formation, i.e., 10\%, being outside the range measured by the ejection age of OBJ-9 to the present day. However, if this fraction is larger, then our estimate of $\epsilon_{\rm ff}$ should be regarded as an upper limit.

%Taking the dynamical timescale as twice the free-fall timescale ($t_{dyn}$ = $2t_{\rm ff}$), the new measurement of 4.68 Myr would mean the formation time ($t_{form}$), i.e., the time over which 90\% of the stars in a cluster form, $\ge$ 7.52$t_{dyn}$, therefore $t_{form}$ $\ge$ 15.04$t_{\rm ff}$. 

%Star formation efficiency per free-fall time ($\epsilon_{\rm ff}$) can be estimated from $\epsilon_{\rm ff} = 0.9\epsilon_* \times t_{\rm ff}/t_{form}$. A calculation from \cite{dario14} yielded $\epsilon_{\rm ff}$ $\sim0.05$, with $t_{form}$ of 5-8 times $t_{\rm ff}$. Therefore, using our $t_{form}$ $\ge$ 15.04$t_{\rm ff}$, the $\epsilon_{\rm ff}$ would be approximately 0.026, even smaller than that of \cite{dario14}, supporting the slow star formation scenarios they proposed.

\section{Summary \& Conclusions}

We have presented follow-up spectroscopic observations of 27 high-priority runaway star candidates from the ONC, based on the 2D (proper motion) traceback analysis of \cite{farias20}, and with the targets selected as showing some indicators of youth, i.e., IR excess and/or variability. The targets were also selected to have relatively old ejection ages, which would place new constraints on the star formation history of the ONC.
%These candidates were pre-defined as possible YSOs by flags developed in \cite{farias20}. 
The primary objective of our work has been to confirm whether these targets are indeed YSOs, primarily by the presense of Li, and, by RV measurement, further confirm an origin in the ONC via 3D traceback.
%more importantly, whether they were ejected from the ONC. 

The candidates were observed using the Magellan 2 + MIKE spectrograph, providing spectra to identify YSO signatures, such as lithium absorption and H$\alpha$ emission, and allowing for 3D traceback based on the measured radial velocity. A summary of our main results is as follows:

\begin{enumerate}[\hspace{-1cm}(a)]
    \item We identified 11 out of 27 targets that exhibit significant lithium absorption (with 5 of these also showing H$\alpha$ emission), confirming their status as low-mass YSOs. 
    
    \item We are able to traceback these confirmed YSOs in 3D and revealed that 8 of the 11 YSOs have a closest approach consistent with an origin in the ONC ($D_{\rm min,3D} < 2.5 - 3$ pc; see Sect.~\ref{section_traceback}).

    \item We cross-match our confirmed YSO runaway candidates with several recent catalogues of clusters, star forming regions and candidate runaways \citep{prisinzano22,kounkel22,hunt23} to check for alternative possible origins for our candidates other than the ONC. We find that one of our confirmed YSOs, OBJ-14, is more likely to have originated from the $\sigma$ Ori cluster given its 3D trajectory and isochronal age, but our runaway candidates with the longest ejection timescales, OBJ-2, OBJ-6 \& OBJ-9, are more likely to originate from the ONC than any other nearby young cluster, given the much greater rate of ejections expected from the ONC compared to other clusters in the region.

    %$\le$10 pc from the ONC, suggesting a possible ejection from the ONC. In general, uncertainties in parallax and RV result in a large ($\sim$5 pc) error on the 3D closest approach. This makes it challenging to precisely determine where in the cluster a runaway exactly originated, but we find for many of our candidates that closest approaches are well-constrained enough to make their origins in the ONC a strong likelihood. %{\bf LaurentC6: Does not sound correct}
    \item Comparing isochronal ages with ejection ages, we find general consistency in the population, but note that the variation among isochronal estimates is large, indicating potentially large systematic uncertainties. We consider that our good runaway candidates from the ONC have utility in helping to refine and calibrate pre-main sequence models. 
    
%    Further testing the candidates involved comparing the time in the past when the candidates were ejected with the age from the PMS evolutionary track. Several candidates had ejection times larger than their isochronal ages, suggesting they are not from the ONC. However, due to the inherent uncertainties in estimating isochronal ages for individual sources, we cannot rule them out based on this criteria. In any case, we found 5 candidates that satisfy both the closest approach and age estimation to be considered as ONC runaways, including the already identified source, V*1781 Ori.
 
    \item Among the likely runaway candidates, we identified 3 with ejection timescales greater than 4 Myr, with the oldest, OBJ-9, being about 4.7~Myr. Consider previous star formation before the ejection and that star formation in the ONC is still ongoing, this implies that the overall formation time of the ONC is likely to be at least 5~Myr. This corresponds to about 10 free-fall times of the system (evaluated at the half-mass radius), indicating a scenario of relatively slow, quasi-equilibrium star cluster formation \citep{tan06}.
    
    \item The oldest ejection age of the sample of 4.7~Myr allows a new estimate of the likely mean star formation efficiency per free-fall time of $\bar{\epsilon}_{\rm ff}\simeq 0.05$. This is similar to the previous estimate of \citet{dario14}, but is now independent of isochronal age estimates based on pre-main sequence evolutionary tracks. The relatively small value of $\bar{\epsilon}_{\rm ff}$ indicates that star formation has proceeded in a relatively slow and inefficient manner, which likely indicates a role for magnetic fields and/or protostellar outflow feedback in regulating its rate \citep[e.g.,][]{2007ApJ...662..395N}.

    %The sources left out from the likely candidates of ONC runaways 
%    {\hskip 1em}\item Based on the number of observed candidates that we rules out as YSOs due to failure to meet the EW(Li) or EW(H$\alpha$) criteria, we conclude that there is a high contamination rate when only conducting traceback analysis in 2D and determining youth solely from the color-magnitude diagram rather than from spectral signatures (only 10 out of 27 targets showed youth signatures from spectra). Further spectroscopic follow-up observations of candidate runaways identified in 2D \citep[e.g., those of ][]{kounkel22} are needed to verify their runaway status.

\end{enumerate}

\section*{Data Availability}
The data underlying this article will be shared on reasonable request to the corresponding author.
 
\section*{Acknowledgments}

MF acknowledges support from the CASSUM program at Chalmers Univ. of Technology. JA and JCT acknowledge support from ERC Advanced Grant MSTAR. This work has made use of data from the ESA space mission Gaia (http://www.cosmos.esa.int/gaia), processed by the Gaia Data Processing and Analysis Consortium (DPAC, http://www.cosmos.esa.int/web/gaia/dpac/consortium). Funding for DPAC has been provided by national institutions, in particular the institutions participating in the Gaia Multilateral Agreement. This research made use of the Simbad and Vizier catalogue access tools (provided by CDS, Strasbourg, France), Astropy \citep{astr13} and TOPCAT \citep{tayl05}.

\bibliographystyle{mnras}
\bibliography{references2}

\appendix
\section{ID 3018141830356350976 (OBJ-20)}

\begin{figure}
    \centering
    \includegraphics[width=0.5\textwidth,center]{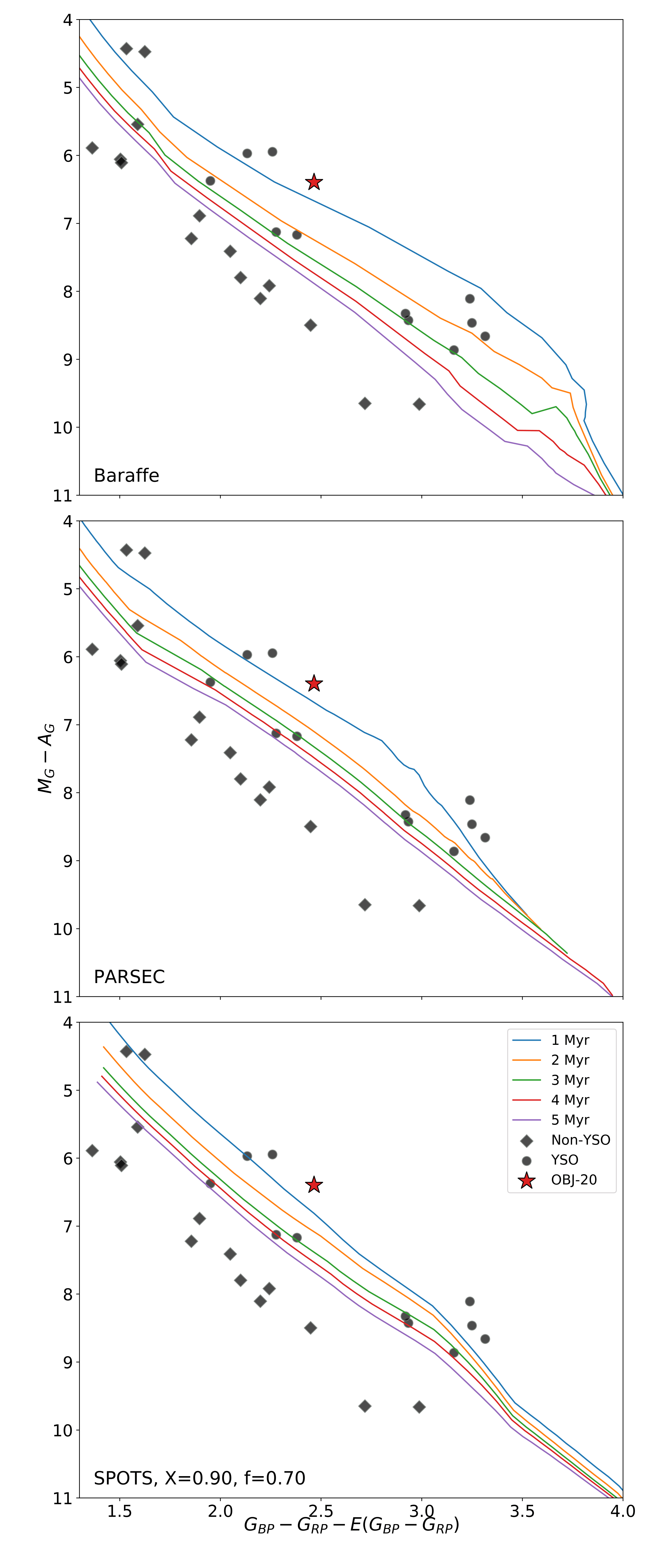}
    \caption{Gaia DR3 BP-RP colour - MG absolute magnitude diagram for OBJ-20 and the rest of our sample including the spectroscopically confirmed YSOs. Extinction AG and reddening E(BP-RP) are estimated per source from Gaia. Overlaid are \protect\citet{baraffe15}, PARSEC \protect\citep{marigo17} and SPOTS \protect\citep[][x=0.9, f=0.7]{somers20} isochrones for 1, 2, 3, 4 and 5 Myr.}
    \label{fig:obj20cmd}
\end{figure}

\begin{figure}
    \centering
    \includegraphics[width=0.5\textwidth]{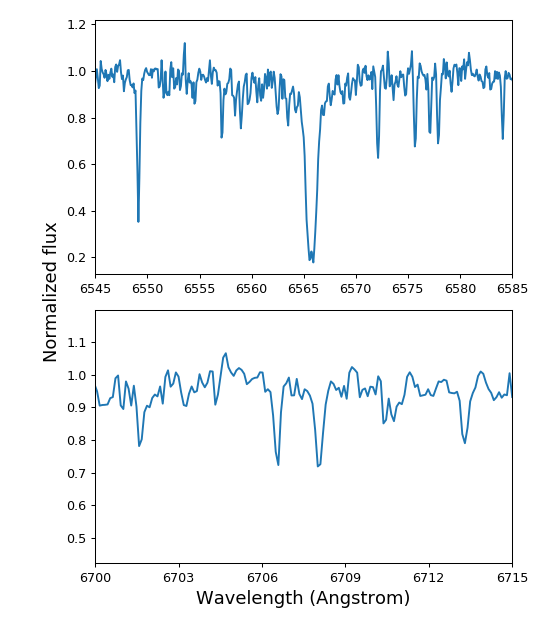}
    \caption{H$\alpha$ (\textit{top)} and lithium 6708 \AA \ (\textit{bottom}) lines for OBJ-20. The fluxes are normalized non-calibrated counts.} %{\bf LaurentC5: Where is the figure 4 referenced in the text?}}
    \label{fig:obj20spectra}
\end{figure}

\bsp
\label{lastpage}	
\end{document}